\documentclass[journal,final,letterpaper,10pt,doublecolumn,twoside]{IEEEtran}

\usepackage{amsfonts}
\usepackage{amssymb}
\usepackage{amsmath}
\usepackage{graphicx}
\usepackage{url}
\usepackage{subfigure} 
\usepackage{cite}
\usepackage{algorithm}
\usepackage{algorithmic}

\begin{document}

\title{Small Cell Traffic Balancing Over Licensed and Unlicensed Bands}

\author{Feilu~Liu,~Erdem~Bala,~
        Elza~Erkip,~Mihaela~C.~Beluri~ and ~ Rui~Yang

\IEEEcompsocitemizethanks{ Copyright (c) 2013 IEEE. Personal use of this material is permitted.
However, permission to use this material for any other purposes must be obtained
from the IEEE by sending a request to pubs-permissions@ieee.org.

 \IEEEcompsocthanksitem Feilu Liu
(feilul@qti.qualcomm.com) is with Qualcomm Technologies, Inc., San
Diego, CA, USA. He was with NYU Polytechnic School of Engineering when this work was done. Elza Erkip (elza@nyu.edu) is
with ECE Dept., NYU Polytechnic School of Engineering,
Brooklyn, NY, USA. Erdem Bala, Rui Yang and Mihaela C. Beluri are
with InterDigital Communications, LLC., Melville, NY, USA. Emails:
\{Erdem.Bala, Mihaela.Beluri, Rui.Yang\}@InterDigital.com. }
\thanks{
The material in this paper was presented in part at the  Third
International Workshop on Indoor and Outdoor Femto Cells (IOFC), May
2011, in Princeton, New Jersey, USA and at the IEEE First
International Workshop on Small Cell Wireless Networks (SmallNets),
June 2012, in Ottawa, Canada. This work is supported by InterDigital
Communications LLC., the New York State Center for Advanced
Technology in Telecommunications (CATT) and the Wireless Internet
Center for Advanced Technology (WICAT), an NSF Industry/University
Cooperative Research Center at NYU Polytechnic School of Engineering. } }

\maketitle

\begin{abstract}
The 3rd Generation Partnership Project (3GPP) recently started standardizing the ``Licensed-Assisted Access using LTE" for small cells, referred to as Dual Band Femtocell (DBF) in this paper, which uses LTE air interface in both licensed and unlicensed bands based on the Long Term Evolution (LTE) carrier aggregation feature. Alternatively,
the Small Cell Forum introduced the Integrated Femto-WiFi (IFW) small cell which simultaneously accesses both the licensed band (via cellular interface) and the unlicensed band (via WiFi interface).  In this paper,
a practical algorithm for IFW and DBF to automatically balance their traffic in licensed and unlicensed bands, based on the real-time   channel,  interference and traffic  conditions  of both bands is described.
The algorithm considers the fact that some ``smart" devices (sDevices) have both cellular and WiFi radios while some WiFi-only devices (wDevices) may only have WiFi radio. In addition, the algorithm considers a realistic scenario where a single small cell user may simultaneously use multiple sDevices and wDevices via either the IFW, or the DBF in conjunction with a Wireless Local Area Network (WLAN).
The goal is to maximize the total user satisfaction/utility of the small cell user, while keeping the interference from small cell to macrocell  below predefined thresholds.
The algorithm can be implemented at the Radio Link Control (RLC) or the network layer of the IFW and DBF small cell base stations.
Results demonstrate that the proposed traffic-balancing algorithm applied to either IFW or DBF significantly increases sum  utility of all macrocell and small cell users,  compared with the current practices. Finally, various implementation issues of IFW and DBF are addressed.
\end{abstract}


\section{Introduction} \label{sec_introduction}
\begin{figure*}
  \center
  \includegraphics[width=6.4in]{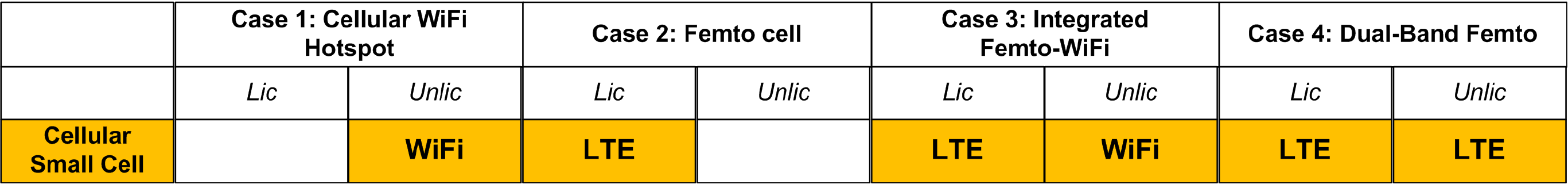}
  \caption{Spectrum and radio access technologies used by each type of small cell. Long Term Evolution (LTE) and WiFi represent the air
interfaces used in a band; blank box means the spectrum is  not used.} \label{table:SpecUse}
\end{figure*}

Small cells as part of the second tier in multi-tiered cellular networks
have been considered as an effective means to boost the capacity and
expand the coverage. Two types of small cells are widely used. One
is the femtocell which shares the cellular licensed band with macrocells
\cite{femto_Survey_JeffAndrews2008,femto-PowerControl-2010Sundeep}.
The other type is the WiFi hotspot that is built by
cellular operators to offload traffic from their licensed bands to
the unlicensed band. Fig.
\ref{table:SpecUse} shows the spectrum map of these two
approaches in Cases 1 and 2, respectively.


%

In this paper, we use the terminology ``device" to refer to the end-user terminal in Long Term Evolution (LTE) and WiFi communications, which is referred to as the user equipment (UE) in 3rd Generation Partnership Project (3GPP) terminology and the ``station" in IEEE 802.11 WiFi terminology.
Today many ``smart" devices such as  smartphones, tablets and iPads are equipped with both WiFi and cellular interfaces. In
order to improve the data rate of such smart devices (sDevices), the Small Cell
Forum proposed the Integrated Femto-WiFi (IFW)
\cite{femto-forum-IFW} which can simultaneously communicate in
both the licensed band (via cellular interface) and the unlicensed
band (via WiFi interface) with sDevices. The IFW spectrum usage is shown in Case 3 of Fig.
\ref{table:SpecUse}.

An alternative way of simultaneously using both the licensed and unlicensed bands is investigated in our earlier study \cite{feilu_femto_accessScheme} which proposes that femto cells can use LTE technology in both licensed and unlicensed bands through the LTE carrier aggregation feature \cite{LTE-A}, resulting in the Dual-Band Femtocell (DBF) in Case 4 of Fig. \ref{table:SpecUse}.  In September 2014, the 3GPP approved the industry proposal \cite{3GPP2014LTE-U} to start standardizing the ``Licensed-Assisted Access using LTE" which is also often referred to as LTE-Unlicensed, LTE-U and U-LTE. The main idea of LTE-U is the same as the DBF framework in this paper.  Since the unlicensed spectrum is shared by many cellular operators and non-cellular devices, how to access the unlicensed band and how to share the unlicensed band with other devices is essential to the DBF user experience. However, these issues have not been addressed in \cite{3GPP2014LTE-U} and may be an important part of the standardization effort.



\begin{figure*}
  \center
  \includegraphics[width=6.4in]{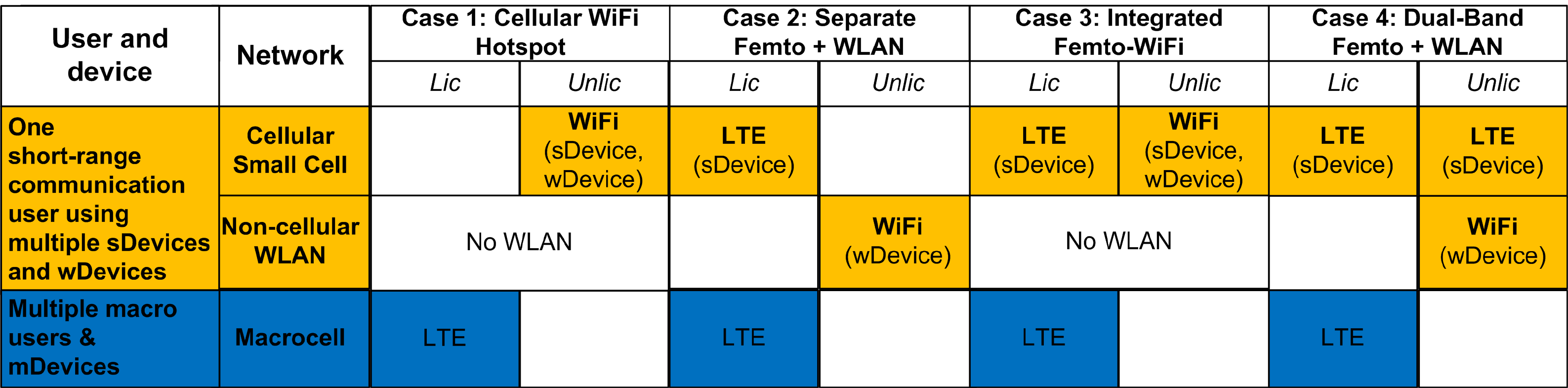}
  \caption{Four use cases considered in this paper. Cases 1 and 2 are the baseline. Cases 3 and 4 are the focus of this paper. LTE and WiFi represent the air interfaces used in a band; blank box means the spectrum is  not used. Note that in Cases 2 and 4, the sDevice can select either the cellular small cell or the non-cellular WLAN; for simplicity, we assume it always selects the cellular small cell.} \label{table:4UseCases}
\end{figure*}

Short-range data communications arising in small cells typically contain different types of devices. One type is the sDevice which is equipped with both WiFi and cellular interfaces as discussed above. We consider LTE as the cellular Radio Access Technology (RAT) in this paper in order to use the LTE carrier aggregation feature for DBF. Another type is the WiFi-only device (wDevice) such as TV, desktop computer, wireless printer and video surveillance camera, which is typically equipped with WiFi but no cellular interface. Cellular-only devices are not considered, as the most recent cellular devices typically have a WiFi interface.
In addition, a single user may use multiple devices at the same time. For example, in a residential scenario, a user may be watching video clips on her tablet jointly over the WiFi and cellular interfaces (using IFW or DBF),  while her wireless video surveillance camera continuously transfers live video to the WiFi access point (AP). Therefore, a user's satisfaction can come from the overall experience from multiple sDevices and wDevices. In Cases 1 and 3, the small cell (WiFi hotspot and IFW) can serve both sDevices and wDevices. However, in Cases 2 and 4, the cellular small cell itself (femto cell and DBF respectively) cannot serve wDevices, hence we assume the femto cell and DBF are deployed with non-cellular wireless local area network (WLAN) APs which are not physically integrated with the femto or DBF base station (BS) in the same box. The four use cases are summarized in Fig. \ref{table:4UseCases}. In the figure and throughout this paper, we denote macro BS and device by mBS and mDevice, respectively, and small cell (of Cases 1, 2, 3 and 4) BS as fBS.

In this study, the ``small cell" mainly refers to the cell for short-range communications in residential and enterprise scenarios, as shown in the four use cases in Fig. \ref{table:4UseCases}; ``macro cell" refers to pico, micro or macro cells. In addition, the ``WiFi" refers to the air interface defined by the IEEE 802.11 standards; the ``WiFi hotspot" only refers to the cellular small cell in Case 1; the ``WLAN" only refers to the non-cellular networks used by the wDevices in  Case 2 and 4.

\begin{figure*}
  \center
  \includegraphics[width=3.4in]{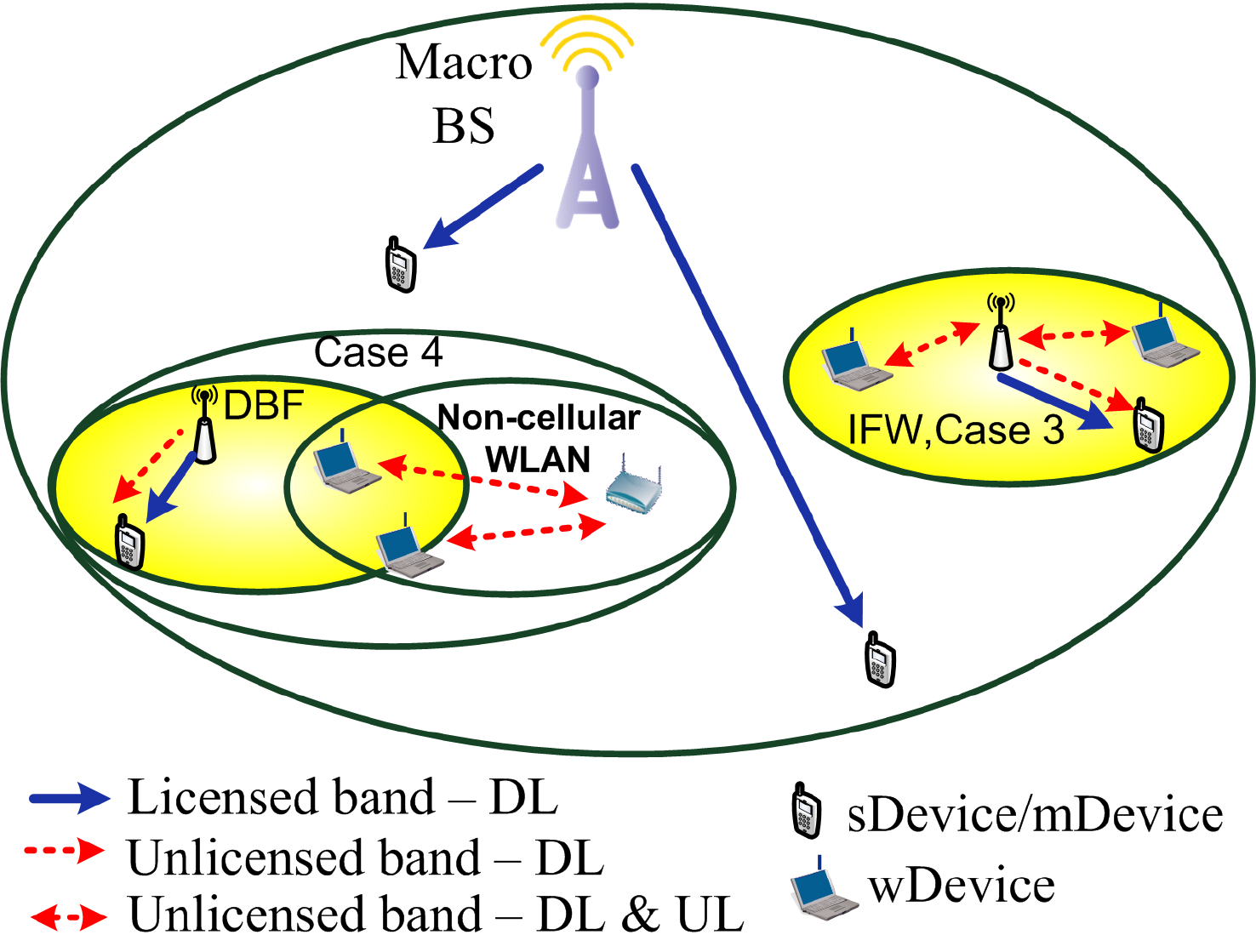}
  \caption{Illustration of the Cases 3 and 4 scenarios considered in this paper. } \label{Fig:coexistence-scenario}
\end{figure*}

The focus of this paper is on the Cases 3 and 4 which is defined in Fig. \ref{table:4UseCases} and illustrated in Fig. \ref{Fig:coexistence-scenario}.  The contribution can be summarized as follows.
\begin{itemize}
  \item  In order for DBF to use the LTE air interface in the unlicensed band, we propose a channel access scheme that aligns with the LTE frame structure. Once the channel is obtained, the DBF will follow the standard LTE air interface in the unlicensed band.

  \item  We propose a dynamic traffic balancing algorithm over licensed and unlicensed bands for IFW and DBF that aims at optimizing the overall \emph{user experience} from multiple sDevices and wDevices in short range communications. The algorithm is based on the real-time channel,  interference and traffic
conditions  of both bands. We formulate and solve for the optimal
downlink traffic balancing scheme in order to maximize the user
utility (satisfaction) from all sDevices and wDevices belonging to the same user while controlling the interference leaked from
the small cell to the macrocell.

\item  The utility maximization described in the previous bullet is achieved by small cell power control in
the licensed band and channel time allocation in the unlicensed
band. Once the optimal channel time usage in the unlicensed band is
determined, the small cell tunes its channel access parameters to
achieve the allocated channel time. The process of tuning channel
access parameters depends on the RAT used
in the unlicensed band. We study how the channel access parameters
can be tuned for the IFW, which uses the WiFi air interface, and for the
DBF, which uses the LTE air interface in the unlicensed band,
respectively.

  \item  We provide extensive system simulations that show that the proposed traffic balancing algorithm significantly improves user satisfaction for IFW and DBF, compared with the current practice where
devices typically have to
choose only one band (licensed or unlicensed) to use at a time, as in Cases 1 and 2 of Fig. \ref{table:SpecUse}.
\end{itemize}


This paper extends our earlier DBF traffic balancing algorithm \cite{feilu_trafficBalancing_conf} by considering multiple non-cellular WLAN devices,  incorporating the IFW scenario, and  introducing a new use case where a single user may use multiple devices. Both this paper and \cite{feilu_trafficBalancing_conf} are based on the channel access scheme proposed in our earlier study \cite{feilu_femto_accessScheme}.
This work is related to \cite{unlicensed-band-LTE} which proposes
that LTE small cells use the licensed-exempt TV whitespace band. It
is proposed in \cite{unlicensed-band-LTE} that LTE small cells use
frequency-hopping and time-hopping in the TV whitespace band to
reduce interference from other devices in the band; whereas this
study proposes a channel-sensing based channel access scheme for LTE
small cells to access the band and reduce interference, which may also be applicable to the new
Study Item (SI) ``Licensed-Assisted Access using LTE"  which was recently approved for 3GPP Rel-13 \cite{3GPP2014LTE-U}. In addition, the existing literature on unlicensed band LTE \cite{unlicensed-band-LTE}\cite{huawei_unlic_LTE}
does not investigate the traffic balancing problem over the two
bands.

Existing traffic balancing algorithms over licensed and unlicensed bands are mainly for IFW \cite{trafficBalance_qos_mag} \cite{fujitsu_trafficBalance_IFW_IncludewifiOnly}, but not for DBF. Specifically, Bennis et. al. \cite{trafficBalance_qos_mag} propose a cross-system learning framework by considering the QoS in traffic balancing and assuming no WiFi-only devices. Elsherif et. al \cite{fujitsu_trafficBalance_IFW_IncludewifiOnly} consider both ``smart" and WiFi-only devices in traffic balancing, with the goal of maximizing the total throughput, but it does not tackle the problem from the perspective of user experience.

The paper is organized as follows.
In Sec. \ref{sec:systemModel}, we provide the system model.
In Sec. \ref{sec: ch-access-scheme}, we introduce a centralized channel access scheme for DBF to use the unlicensed band.
In Sec. \ref{sec:DL-assignment}, we propose a traffic balancing algorithm for small cells to assign traffic over the licensed and unlicensed bands.
The RAT-dependent process of tuning channel access parameters is analyzed in Sec. \ref{sec_IFW} for IFW and in Sec. \ref{sec_DBF} for DBF. In Sec. \ref{sec:Femto-system-simulations}, we
evaluate the proposed traffic balancing algorithm through system simulations. In Sec. \ref{sec:conclusion}, we conclude the paper and compare IFW and DBF from an implementation standpoint.

\section{System Model}\label{sec:systemModel}

Two types of small cells, the IFW
\cite{femto-forum-IFW} introduced by the Small Cell Forum and the DBF
proposed in \cite{feilu_femto_accessScheme}\cite{huawei_unlic_LTE}, that simultaneously access both licensed and
unlicensed bands are considered in this paper.
We consider closed access small cells
that can only be accessed by registered devices
\cite{femto_Survey_JeffAndrews2008}.  In the licensed band, the LTE
air interface \cite{LTE-A}, which divides the spectrum into radio
blocks which are referred to as subchannels, is used. In the unlicensed band, different nodes share the air resource in time, not frequency, so we do not consider subchannels.

Throughout this
paper, ``WiFi hotspot," ``IFW" and ``DBF" refer to both the fBS and
all associated devices using the appropriate radio access technology. The term ``WLAN" refers to the network formed by a WiFi AP and wDevices that coexist with the sDevices in Cases 2 and 4; while the term ``WiFi hotspot" refers to the small cell in Case 1 of Fig. \ref{table:SpecUse} that is used by both sDevices and wDevices.

We assume that IFW
and DBF BSs conduct the traffic balancing over the
licensed and unlicensed bands.
How to allocate radio resources
(i.e., power, frequency and time) to individual \emph{devices within
one cell} in the licensed band is a complex problem \cite{ofdm_multiuser}, and is out of
the scope of this paper where the focus is radio resource allocation
\emph{among different types of cells} including macrocell, small
cell (IFW or DBF) and non-cellular WLAN. Therefore, for simplicity,
we only consider a single sDevice in the small cell; the extension to
the multi-device small cell case could be based on an analysis similar
to this paper. In addition, we assume that the licensed band and unlicensed band use separate power budgets, due to different government regulation requirements for the bands.

In the IFW use case, we consider an IFW fBS and a mBS, where the IFW fBS is connected to one sDevice and $N_W$ wDevices and the mBS is serving $N_M$ mDevices. Whereas in the DBF use case, we consider a DBF fBS, a WiFi AP and a mBS, where the DBF fBS is connected to one sDevice, the WiFi AP is connected to $N_W$ wDevices, and the mBS is serving $N_M$ mDevices. The WiFi AP may or may not be physically integrated with the DBF fBS. We consider the case where the WiFi WLAN and DBF use the same unlicensed band carrier frequency,\footnote{Though the bandwidth of all unlicensed band is large, typically a device only supports a limited number of the bands to lower the device cost. In dense WiFi and small cell deployments in locations such as enterprise and urban residential apartment buildings, where high interference can be observed on many unlicensed frequencies, some near-by WLANs or DBF small cells may have to use the same unlicensed band carrier frequency.} which is the worst case in terms of network performance. In either use case, the sDevice and wDevices are used by a single user or a single group of users (e.g., a family, an enterprise or passengers on the same vehicle). In the unlicensed band, the DBF fBS contends with $N_W$ wDevices for the channel. We assume that $N_W$ is much larger than one, and each contending node (wDevice or DBF fBS) can sense the other nodes.  Furthermore, the DBF fBS success probability for each access attempt in the unlicensed band is denoted as $P_{\textmd{DBFsuc}}$. If $T_{\textmd{attempt}}$ (the time interval between two channel access attempts from the DBF fBS) is comparable to the transmission durations of the other unlicensed band devices, it is reasonable to assume that the fBS channel access attempts are statistically independent.

We assume no external interference in the unlicensed band except the collisions among the transmitters in the IFW, DBF and WLAN; in case of collisions, we assume that WiFi transmissions always fail, otherwise, transmission errors are neglected since the data rate is adapted to the instantaneous SINR \cite{802.11-2007}. In addition, hidden terminal and exposed terminal problems can be detected by the DBF and IFW fBSs via existing LTE downlink channel quality indicator (CQI) feedbacks over the licensed band. In the unlicensed band, if the fBS senses good channel quality while the CQI from the UE report is constantly below a threshold, the fBS may determine that the UE is under high interference from hidden terminals. Similarly, if the fBS senses bad channel quality while the CQI from the UE report is constantly above a threshold, the fBS may determine that the fBS itself experiences exposed terminal problems. Note that the detection of hidden and exposed terminals is difficult, if not impossible, when only one unlicensed carrier frequency is used (e.g., WiFi, Bluetooth).
The fBS may take different approaches upon detecting the hidden terminals, e.g., selecting another unlicensed carrier frequency to operate on, which is out of the scope of this paper.

The LTE air interface \cite{LTE-A} supports both Frequency-Division Duplex (FDD) and Time-Division Duplex (TDD) modes.
We consider FDD-mode LTE in this study. For ease of exposure, we only consider downlink transmissions for mDevices and sDevices, and assume that sDevices use the unlicensed band only for downlink transmissions in both IFW and DBF use cases (uplink is in the licensed band). For the wDevices, we consider both downlink and uplink transmissions, since the downlink contends with uplink in a random fashion and cannot be separately studied.
In the unlicensed band, the performance of the sDevices in both IFW and DBF use cases is dependent on the traffic load of the coexisting wDevices,
which can be described by the parameter $\bar{t}_w$, the fraction of
channel time needed to deliver the UL and DL traffic of all wDevices.
In general, $\bar{t}_w$ is determined by the  average traffic load
and data rate of every wDevice. Note that the UL and DL data rates of a wDevice are the same due to channel reciprocity.	 We define a wDevice’s throughput as the sum of uplink and downlink throughput. In order to identify the maximum capacity of the two-tiered cellular
network, we assume that the mDevices and sDevices always have downlink traffic to receive (i.e., their traffic loads are more than their physical layers can support).   In the licensed band, we assume that mBSs do not adjust their transmission powers
in the presence of small cell interference.

Both LTE and WiFi have multiple modulation and coding schemes (MCSs)
and adapt their MCSs to the instantaneous
signal-to-interference-plus-noise ratios (SINRs). In practice, the
WiFi rate function $R_W(\cdot)$ and LTE rate function $R_L(\cdot)$
are dependent on their MCSs and bandwidth. In this paper, we will
consider the actual $R_W(\cdot)$ determined by the WiFi standard
\cite{802.11-2007}. For ease of exposure, we will first conduct our
analysis using Shannon capacity as $R_L(\cdot)$ in Section
\ref{sec:opt_solution}; then in Section
\ref{subsec:Traffic-assign-Implementation}, we will consider a
closed-form approximation for the actual LTE rate function.   It is
clear that the small cell unlicensed band rate function $R_U(\cdot)$
is equal to $R_W(\cdot)$ and $R_L(\cdot)$ in IFW and DBF,
respectively. We assume that the fBS knows the received SINRs at the
sDevice in both licensed and unlicensed bands via device feedback.
More specifically, in the licensed band, the fBS controls the transmission power $P_f^{(k)}$ and knows the received SINR $P_f^{(k)} \gamma_f^{(k)}$ of sDevice in subchannel $k$ ($k=1,2,\dots,K$). Here in subchannel $k$, $\gamma_f^{(k)}$  is  path loss of the desired signal divided by the interference and noise power.
We also assume that the fBS knows the inter-cell
interference channel gain $h_{fm}^{(k)}$ ($k=1,2,\dots,K$) from the
fBS to the mDevice that uses subchannel $k$ in the licensed band. For
simplicity, we do not consider fading, mobility or multi-antenna
transceivers, which would mainly affect the required overhead for
obtaining the SINRs and channel gains in our problem formulation.

A utility function $U(S)$ is used to evaluate user satisfaction about an achieved throughput $S$. We will consider the widely-used logarithmic  utility function  to achieve proportional fairness \cite{log_utility},
\begin{eqnarray}
U(S) &=& \ln(S), \label{eq:utilityfunction}
\end{eqnarray}
where $\ln(\cdot)$ is natural logarithm function. The concavity of the logarithmic  function well captures the typical user experience about throughput -- as throughput increases, user satisfaction (utility) grows faster when throughput is low than when it is high.

\section{A DBF Channel Access Scheme for Unlicensed-Band LTE}\label{sec: ch-access-scheme}
The LTE-Advanced standard \cite{LTE-A} introduces the carrier aggregation feature, which allows up to five component carriers (CCs) to be aggregated to form a single LTE radio interface with a bandwidth of up to 100MHz in both downlink and uplink. The CCs can be either contiguous, non-contiguous or in different bands \cite{LTE-A}.
Our proposed DBF uses the LTE air interface in both licensed and unlicensed bands via the LTE carrier aggregation feature.

LTE was designed based on the assumption of exclusive spectrum use,
which is not true in the unlicensed band where devices with
different air interfaces coexist. However, existing channel access
schemes in the unlicensed band such as the IEEE 802.11 distributed
coordination function (DCF) and point coordination function (PCF)
\cite{802.11-2007}, are not designed for cellular air interfaces,
and do not align with the LTE frame structure. LTE transmissions are
organized in periodic subframes in time, and can start only at the
beginning of subframes \cite{LTE-A}. As a result, channel access
attempts in the unlicensed band must take place right before the
start time of subframes. Otherwise, even though the fBS obtains the
channel, it cannot transmit until the start time of the next
subframe, and may lose the transmission opportunity since other
unlicensed-band devices will find the channel idle and transmit.
Therefore, in this section, we propose a channel access scheme that
aligns with LTE frame structure. Once the access to the unlicensed
band is obtained, the fBS will follow the standard LTE air interface and
assign radio resources to the sDevices through the licensed-band control
channel.

Two guidelines are followed in the design of the DBF channel access scheme for the unlicensed band: 1) The fBS senses the unlicensed spectrum in order to avoid interference from ongoing transmissions by other unlicensed-band devices.  2) The channel access scheme aligns with LTE frame structure.

\begin{figure}
  \center
  \includegraphics[width=3.1in]{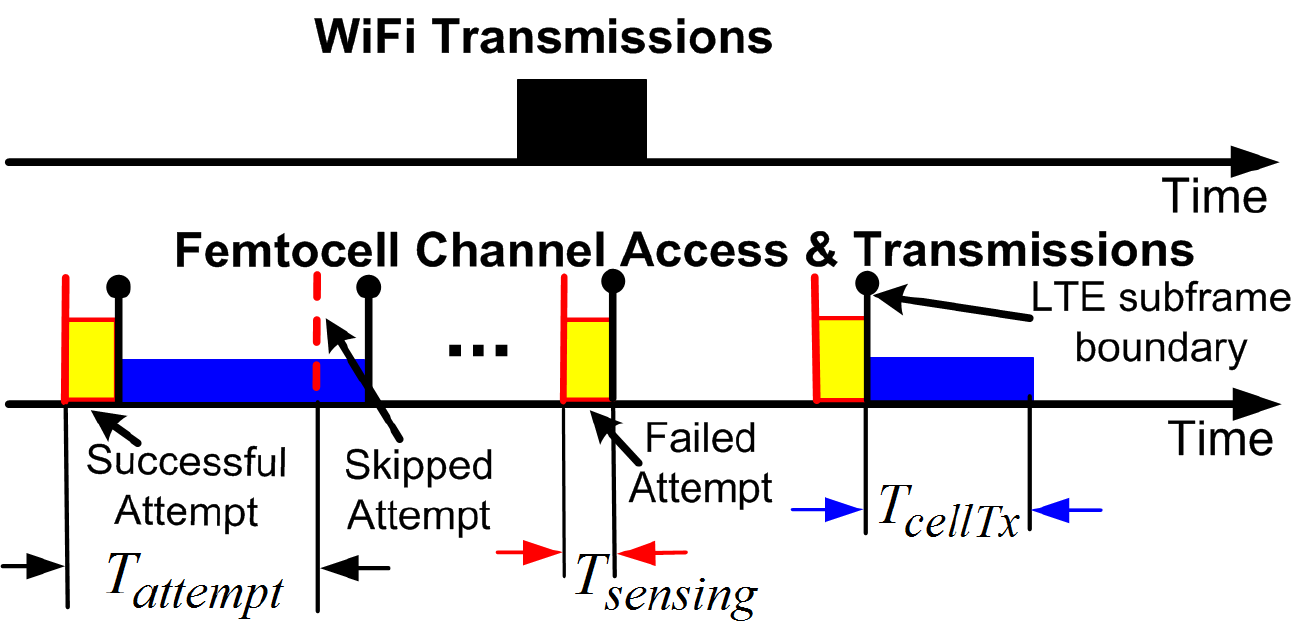}
  \caption{Dual-Band Femtocell (DBF) channel access mechanism in the unlicensed band.} \label{fig_channel_access}
\end{figure}

Fig. \ref{fig_channel_access} illustrates the proposed channel
access scheme. The fBS attempts to access the channel at
pre-assigned periodic time instants, called ``access opportunities."
The period of the access opportunities is denoted as
$T_{\textmd{attempt}}$. At each access opportunity,
the fBS senses the unlicensed band, which takes $T_\textmd{sensing}$ seconds.
If the channel is idle, the fBS
will access the channel and use it for a fixed duration,
$T_{\textmd{cellTx}}$; otherwise, the fBS will wait for the next
access opportunity.

As shown in Fig. \ref{fig_channel_access}, to fit this channel access scheme with the periodic LTE subframe structure, we require both $T_{\textmd{attempt}}$ and $T_{\textmd{cellTx}}$ should be integer multiples of LTE subframe duration which is 1ms \cite{LTE-A}. Also, the $T_{\textmd{attempt}}$ includes the $T_\textmd{sensing}$ and the access opportunity must be $T_\textmd{sensing}$ before the LTE subframe boundary so that the fBS
can complete sensing the unlicensed band right at the LTE subframe boundary and transmit using the whole LTE subframe.   Moreover,
As we will see in Sec. \ref{sec_DBF}, a DBF fBS
can adjust its unlicensed band usage by tuning the
parameters $T_{\textmd{attempt}}$ and $T_{\textmd{cellTx}}$.
In practice, the channel sensing time $T_\textmd{sensing}$  is mainly determined by the hardware and is on the order of
10 microseconds \cite{802.11-2007}  which is far less than the $T_{\textmd{attempt}}$ and $T_{\textmd{cellTx}}$, therefore having negligible impact to DBF performance.

In order to prevent DBFs from keeping the channel for a long time,
the fBS should not access the channel immediately after a channel
use. 
If the end of a transmission happens to be an access
opportunity, the fBS should skip it; if the end of a
transmission is in between two access opportunities, the fBS should skip
the access opportunity immediately following the end of the transmission. This guarantees that
the DBF leaves at least $T_{\textmd{attempt}}$ seconds between
two consecutive transmissions for other coexisting devices to access the
unlicensed band.

%


\section{Small Cell Traffic Balancing Over Licensed and Unlicensed Bands}\label{sec:DL-assignment}

In this section, we formulate a traffic balancing strategy for dual-band small cells in Cases 3 and 4 of Fig. \ref{table:4UseCases} to assign traffic over the licensed and unlicensed bands. The formulation is independent of the unlicensed band RAT, hence applicable to both IFW and DBF; whereas the implementation will be RAT-dependent and will be described in Sec. \ref{sec_IFW} for IFW and in Sec. \ref{sec_DBF} for DBF.

\subsection{Transmission Parameters for Traffic Balancing}\label{subsec:Parameters-To-Optimize}

The IFW and DBF access the unlicensed band based on channel sensing,
so at most one device can use the channel at any given time, except
for collisions. Hence, the unlicensed band is shared in time among
different devices, and the unlicensed-band usage can be best
characterized by the fraction of time that a device occupies the
channel. We will control  small cell unlicensed-band usage by
tuning its fraction of channel time $t_f$, which will impact $t_w$,
the total fraction of channel time used by all the wDevices. The licensed
band is simultaneously utilized by all small and macro cells, and
some mDevices may experience severe interference from small cells
\cite{femto-PowerControl-2010Sundeep}. We will adjust fBS
transmission power $P_f^{(k)}$ in subchannel $k$, so that the
interference to mDevices can be controlled, while the desired
performance for sDevices is obtained.

\subsection{Downlink User Utility Optimization for DBF and IFW Use Cases}\label{sec:user utility opt}

Recall that in the system model in Section \ref{sec:systemModel},
for both DBF and IFW small cells, there is an sDevice and $N_W$ wDevices which are used by a single user or single group of users. The sDevice shares the unlicensed band with $N_W$ wDevices and the licensed band with $N_M$ mDevices.
The buffer status (e.g., full-buffer or not) of the wDevices depends on not only their aggregate load $\bar{t}_w$, but also the DBF or IFW cell's channel time usage $t_f$.   For example, if $\bar{t}_w=50\%$, the wDevices will not be in full-buffer status when there is no other unlicensed band user; however, in the DBF and IFW scenarios where $t_f=60\%$,  the wDevices will be in full-buffer status. Recall that the sDevices  always have traffic to receive, so the optimal $t_f$ should be such that
\begin{equation}
t_{max}-t_f \leq \bar{t}_w \label{eq:tmax-tf<twbar}
\end{equation}
where $t_{max}$ is
the maximum fraction of time that the unlicensed
band can be used;\footnote{The $t_{max}$ is determined by the channel
access schemes in the unlicensed band, and is strictly less than one
if channel-sensing based access schemes are used.  If
$t_{max}=1$, then the unlicensed-band channel is always being used
and all devices will detect channel busy and not access the channel,
which contradicts the assumption $t_{max}=1$.  } otherwise, part of the available unlicensed band channel will be unused, resulting in suboptimality. Consequently, with optimal traffic
balancing in the DBF or IFW cells, the wDevices will always be in full buffer status \cite{queuing-theory-textbook}, although their traffic loads may be limited.

It has been shown in
\cite{Bianchi,WiFi_modeling_Foh_freezingCounter,WiFi_analysis_freezingCounter_Lee2005Conf}  that for a WLAN with full-buffer stations and no other unlicensed band users,  WLAN station $i$'s fraction of channel time $\alpha_i$ is determined by the channel access parameters and channel conditions of all stations. Hence, in DBF or IFW use case with optimal traffic  balancing,  the transmission time $t_{w,i}$ of wDevice $i$ (in terms of fraction of the whole channel time) is
\begin{equation}
t_{w,i} = \alpha_i  t_w
, ~~ i=1,2,\dots,N_W,
\end{equation}
where $t_w$ is the channel usage of
all $N_W$ wDevices.
Then the throughput of a wDevice is
\begin{equation}
S_{W,i} = R_{W,i} t_{w,i}  = R_{W,i}\alpha_i  t_w  , ~ i=1,2,\dots,N_W ,   \label{eq:R_w}
\end{equation}
where $R_{W,i}$ is determined by WiFi rate function $R_W(\cdot)$ and
device $i$'s  instantaneous SINR. In Section \ref{sec:opt_solution}, we show that  $\alpha_i$ has no impact to the final optimal solution.

The  throughput of the  sDevice  is from both the licensed and unlicensed bands. In the licensed band,  the sDevice and mDevices simultaneously use the band, so power control is needed to keep the interference from the fBS to the mDevices below given thresholds. In the unlicensed band, the ``listen before talk" style of channel access is widely used (e.g., WiFi), so interference is not a major issue, hence we do not apply power control.  Recall that the fBS knows downlink SINR via device feedback, in addition, the licensed and unlicensed bands use separate power budgets due to different government regulation requirements, so the unlicensed band data rate $R_U$  is a constant that is determined by WiFi rate function $R_W(\cdot)$ (for IFW), LTE rate function $R_L(\cdot)$ (for DBF) and
 the instantaneous SINR in the unlicensed band. The unlicensed band channel is shared by sDevices and wDevices in time, so we control the fraction of channel time $t_f$ that is used by the sDevice. Then the sDevice throughput is
\begin{eqnarray}
S_f &=& \sum_k R_L(P_f^{(k)} \gamma_f^{(k)}) + t_f R_U  , \label{eq:R_f}
\end{eqnarray}
where $P_f^{(k)}$ is the transmission power in subchannel $k$ of the
licensed band.



The optimization problem can be formulated as,
\begin{eqnarray}
\max_{P_f^{(k)}, ~ t_f, ~t_w} &&
U_{sum} = \sum_{i=1}^{N_W} U( R_{W,i}\alpha_i  t_w  )  ~ +  \nonumber  \\
&& U \left (\sum_k R_L(P_f^{(k)} \gamma_f^{(k)}) + t_f R_U \right ),  \label{opt_prob:obj}\\
\textmd{Subject to} && P_f^{(k)} |h_{fm}^{(k)}|^2  \leq \bar{I}_k, ~ k = 1, 2, \dots, K , \label{opt_prob:cond_Imax} \\
&& t_f + t_w \leq  t_{max} , \label{opt_prob:range_tf} \\
&& t_w \leq \bar{t}_w  \label{opt_prob:range_tbar} \\
&& \sum_{k} P_f^{(k)} \leq P_{tot} , \label{opt_prob:cond_Ptot} \\
&& t_f \geq 0, t_w \geq 0,   P_f^{(k)} \geq 0, ~ k = 1, 2, \dots, K
. \nonumber \\  \label{opt_prob:range_Pf}
\end{eqnarray}
Constraint (\ref{opt_prob:cond_Imax}) follows the widely-adopted
principle \cite{femto-PowerControl-2010Sundeep} in two-tiered
networks which requires that the interference power leaked from the
small cell to the macrocell cannot exceed the maximum allowed
interference temperature  $\bar{I}_k$ in subchannel $k$ ($k = 1, 2,
\dots, K$), which are predefined system parameters that determine
the performance tradeoff between macro and small cells in the
licensed band. Constraint (\ref{opt_prob:range_tf}) shows the fact
that, in practice, the total unlicensed band usage must be less or
equal to the maximum fraction of time $t_{max}$ that the unlicensed
band can be used.  Constraint
(\ref{opt_prob:range_tbar}) specifies that the aggregate wDevice channel usage cannot
exceed the time determined by their aggregate traffic load.

For the convenience of presentation, we treat $t_w$ as a variable for the fBS to optimize in the problem formulation above;  however, the final solution in Section \ref{sec:opt_solution} shows that the fBS does not need to adjust $t_w$ -- whenever the fBS adjusts its $t_f$,  the $t_w$ is automatically adjusted.
The objective (\ref{opt_prob:obj}) maximizes the total user experience/utility from all the sDevice and wDevices used by the user (or group of users), and is equivalent to 
\begin{equation}
\max_{P_f^{(k)}, ~ t_f, ~t_w} ~  U \left (\sum_k R_L(P_f^{(k)}
\gamma_f^{(k)}) + t_f R_U \right )  +  N_W U(t_w ) .
\label{opt_prob:obj-simplified}
\end{equation}

The mathematical formulation (\ref{opt_prob:obj})-(\ref{opt_prob:range_Pf}) shares some similarities with the optimization problem in \cite{femto-PowerControl-2010Sundeep}; however, unlike \cite{femto-PowerControl-2010Sundeep} which only considers the licensed band, our optimization considers traffic balancing over both licensed and unlicensed bands.

\subsection{Solution to the Optimization Problem} \label{sec:opt_solution}
In the above optimization problem, $P_f^{(k)}$ affects the sum
utility (\ref{opt_prob:obj}) only through the  small cell
licensed-band throughput  $\sum_k R_L(P_f^{(k)} \gamma_f^{(k)})$. In
addition,  the utility function $U(S)=\ln(S)$ is strictly increasing
with throughput $S$, hence maximizing  small cell  licensed-band
throughput $\sum_k R_L(P_f^{(k)} \gamma_f^{(k)})$  subject to
(\ref{opt_prob:cond_Imax}) and (\ref{opt_prob:cond_Ptot}) will
optimize the sum utility (\ref{opt_prob:obj}) as well. Therefore, to
find the optimal $P_f^{(k)}$, we solve the following optimization
problem,
\begin{eqnarray}
\max_{P_f^{(k)}} && \sum_k R_L(P_f^{(k)} \gamma_f^{(k)}) \label{opt_prob2:obj} \\
\textmd{Subject to} && \sum_{k} P_f^{(k)} \leq P_{tot} , \label{opt_prob2:cond_Ptot} \\
&& 0 \leq P_f^{(k)}  \leq \bar{I}_k / |h_{fm}^{(k)}|^2 , ~ k = 1, 2, \dots, K . \nonumber \\
&&  \label{opt_prob2:cond_Pmax_k}
\end{eqnarray}
After obtaining the optimal $P_f^{*(k)}$, finding the optimal $t_f$ to the original optimization problem (\ref{opt_prob:obj})-(\ref{opt_prob:range_Pf}) is equivalent to the following,
\begin{eqnarray}
\max_{t_f, ~t_w} && U \left ( \sum_k R_L(P_f^{*(k)} \gamma_f^{(k)}) + t_f R_U \right  )  + \nonumber \\
&& N_W U( t_w ) , \label{opt_prob3:obj} \\
\textmd{Subject to}
&& t_f + t_w \leq  t_{max} , \label{opt_prob3:tmax} \\
&& t_f \geq 0, ~~ 0 \leq t_w \leq \bar{t}_w  . \label{opt_prob3:range_tf}
\end{eqnarray}
We first consider the optimization in (\ref{opt_prob2:obj})-(\ref{opt_prob2:cond_Pmax_k}).  In this subsection we solve the problem assuming that the rate function $R_L(\cdot)$ is given by Shannon capacity, that is
\begin{eqnarray}
R_L(P_f^{(k)} \gamma_f^{(k)}) = B \log_2(1+P_f^{(k)} \gamma_f^{(k)}), \label{eq:ratefunction}
\end{eqnarray}
where $B$ is the bandwidth of a small cell subchannel in the licensed band.
In Section \ref{subsec:Traffic-assign-Implementation} we will discuss how this analysis can be extended to the case when $R_L(\cdot)$  is obtained using an approximation to the LTE rate function.
Using the Karush-Kuhn-Tucker (KKT) conditions \cite{ConvexOptimization_textbook},   it is easy to see that the solution to (\ref{opt_prob2:obj})-(\ref{opt_prob2:cond_Pmax_k}) with $R_L(\cdot)$ defined in (\ref{eq:ratefunction}) is given by
\begin{eqnarray}
 P_f^{*(k)} =  \min \left( \left(\frac{1}{\mu}  - \frac{1}{\gamma_f^{(k)}}   \right)^+,  ~\frac{\bar{I}_k}{|h_{fm}^{(k)}|^2} \right)  , \label{opt_prob:Pf_expr}
\end{eqnarray}
Here $x^+ = \max(0,x)$, and $\mu$ is chosen to satisfy (\ref{opt_prob2:cond_Ptot}) with equality.
The solution in (\ref{opt_prob:Pf_expr}) can be numerically obtained by the \emph{modified water-filling algorithm} \cite{cap-limited-water-filling-wcnc2009}\cite{cap-limited-water-filling-EURASIP2008},  which allocates power into subchannels similar to regular water-filling procedure, with the only difference that the power in subchannel $k$ must be below $\bar{I}_k / |h_{fm}^{(k)}|^2$.

Based on the optimal $P_f^{*(k)}$ obtained above, we next solve the second optimization problem (\ref{opt_prob3:obj})-(\ref{opt_prob3:range_tf}).  Since $P_f^{*(k)}$ has been determined, the total LTE licensed band rate of the sDevice,
\begin{equation}
R^{\emph{tot}}_L = \sum_k R_L(P_f^{*(k)} \gamma_f^{(k)}),
\label{opt_prob:obj-RL-sum}
\end{equation}
is now a constant. The objective function (\ref{opt_prob3:obj}) is an increasing function of $t_f$ and $t_w$ (recall that $U(\cdot) = \ln(\cdot)$), so equality must be achieved in Constraint (\ref{opt_prob3:tmax}) to maximize (\ref{opt_prob3:obj}), hence we have
\begin{equation}
t_w=t_{max} - t_f. \label{opt_prob3:tmax-equality}
\end{equation}
Submit (\ref{opt_prob:obj-RL-sum}) and (\ref{opt_prob3:tmax-equality}) into the optimization problem (\ref{opt_prob3:obj})-(\ref{opt_prob3:range_tf}), we obtain the simplified formulation below,
\begin{eqnarray}
\max_{t_f} && \ln ( t_f + R^{\emph{tot}}_L / R_U )  +  N_W \ln(t_{max} - t_f ) \label{opt_prob4:obj} \\
\textmd{Subject to}
&& t_f \geq 0, ~~ 0 \leq t_w \leq \bar{t}_w  . \label{opt_prob4:range_tf}
\end{eqnarray}
We temporarily ignore the Constraint (\ref{opt_prob4:range_tf}), take the derivative of (\ref{opt_prob4:obj}) with respect to $t_f$ and set the derivative to zero, we can see that
\begin{equation}
t_f^* =  \frac{1}{N_W +1}  \left( t_{max} - N_W
\frac{R^{\emph{tot}}_L}{R_U}  \right).
\end{equation}
Then we consider the Constraint (\ref{opt_prob4:range_tf}) which defines the fixed traffic load of the wDevices, from which we have
\begin{equation}
t_f^* \geq   (t_{max} - \bar{t}_w)^+,
\end{equation}
because sDevices always have data to receive; otherwise,  part of the available unlicensed band channel will be unused which results in suboptimal $t_f^*$.
Therefore,
\begin{eqnarray}\nonumber
t_f^* &=& \max \left(  \frac{1}{N_W +1}  \left( t_{max} - N_W
\frac{R^{\emph{tot}}_L}{R_U}
   \right)^+ ,   \right. \\
&&\left.   ~~~~~~~~ (t_{max} - \bar{t}_w)^+   ~ \right)
\nonumber \\
 &=&  \max \left(   ~~~ (t_{max} - \bar{t}_w)^+   ~ \right. , \nonumber \\
&& \left.  \frac{1}{N_W +1}  \left( t_{max} - N_W
\frac{\sum_k R_L(P_f^{*(k)} \gamma_f^{(k)}) }{R_U}
   \right)^+    \right) \nonumber \\
   &&  \label{opt_prob:tf_expr} \\
t_w^*&=& t_{max} - t_f^* .  \label{opt_prob:tw_expr}
\end{eqnarray}
The solution (\ref{opt_prob:tw_expr}) shows that  the fBS can control the wDevice channel usage $t_w$ by adjusting the sDevice channel usage $t_f$.

The fBS will compute the optimal transmit power $P_f^{*(k)}$ in the licensed band and the optimal transmission time $t_f^*$ in the unlicensed band using (\ref{opt_prob:Pf_expr}) and (\ref{opt_prob:tf_expr}), respectively. The fBS will then adjust the amount of traffic assigned to the unlicensed band so that it transmits in the unlicensed band for $t_f^*$ fraction of time. The fBS will assign the rest traffic to the licensed band. In addition, the fBS will transmit at a power of $P_f^{*(k)}$  in subchannel $k$ of the licensed band.  Note that although wDevice data rate $R_W$ appears in our problem formulation (\ref{opt_prob:obj}), it does not appear in the final solutions of  $P_f^{*(k)}$ and $t_f^*$.
As a result, the fBS does not have to obtain wDevice data rate information to carry out the optimal traffic-balancing scheme.

\begin{figure*}[hpbt]
\centering \subfigure[Optimal $t_f$ value (Eq. \ref{opt_prob:tf_expr})]{
        \includegraphics[width=3.0in]{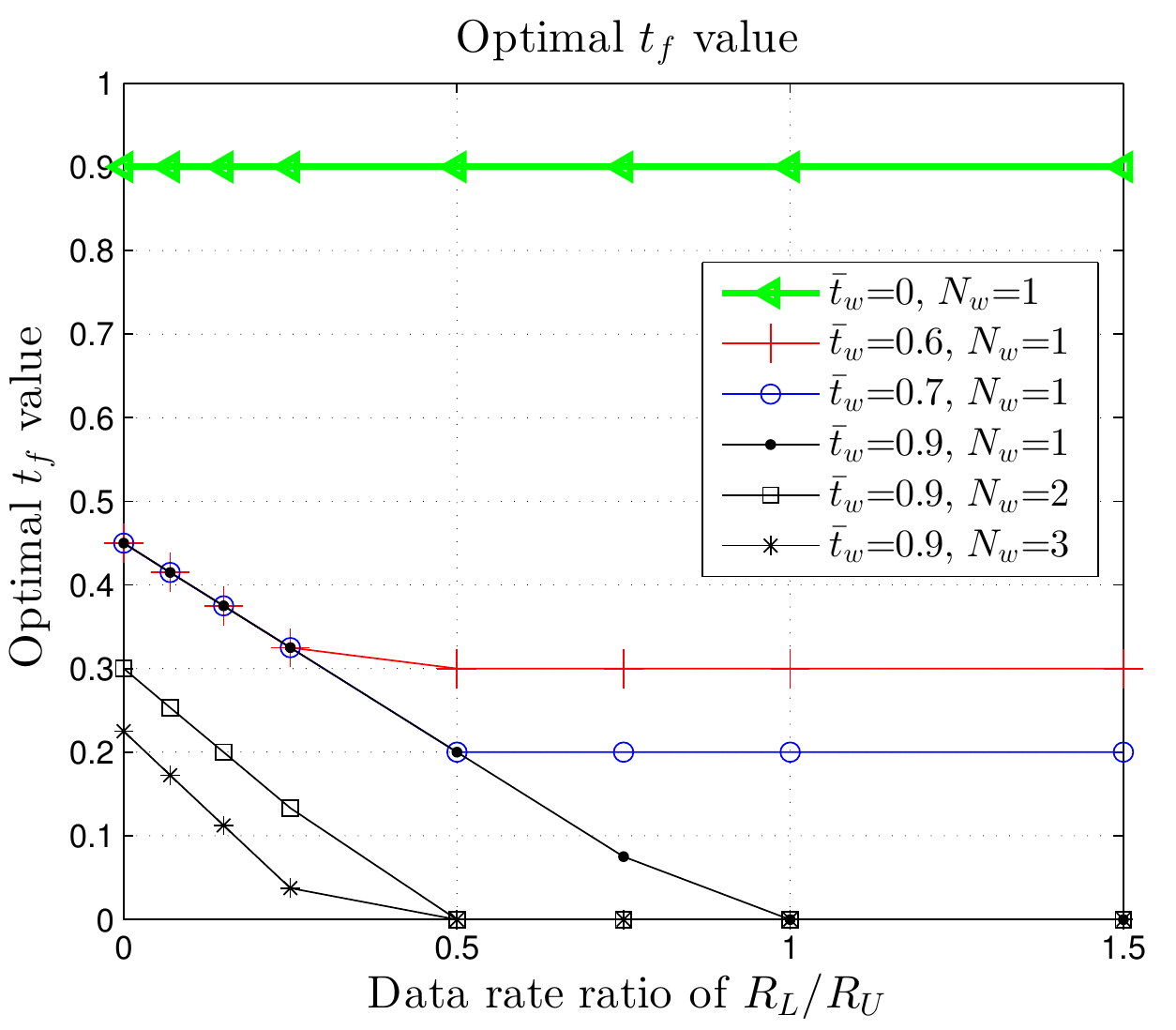}
        \label{fig:plot_tf}
        } \hfill
 \centering \subfigure[Sum utility of sDevices and wDevices (Eq. \ref{opt_prob:obj})]{
        \includegraphics[width=3.13in]{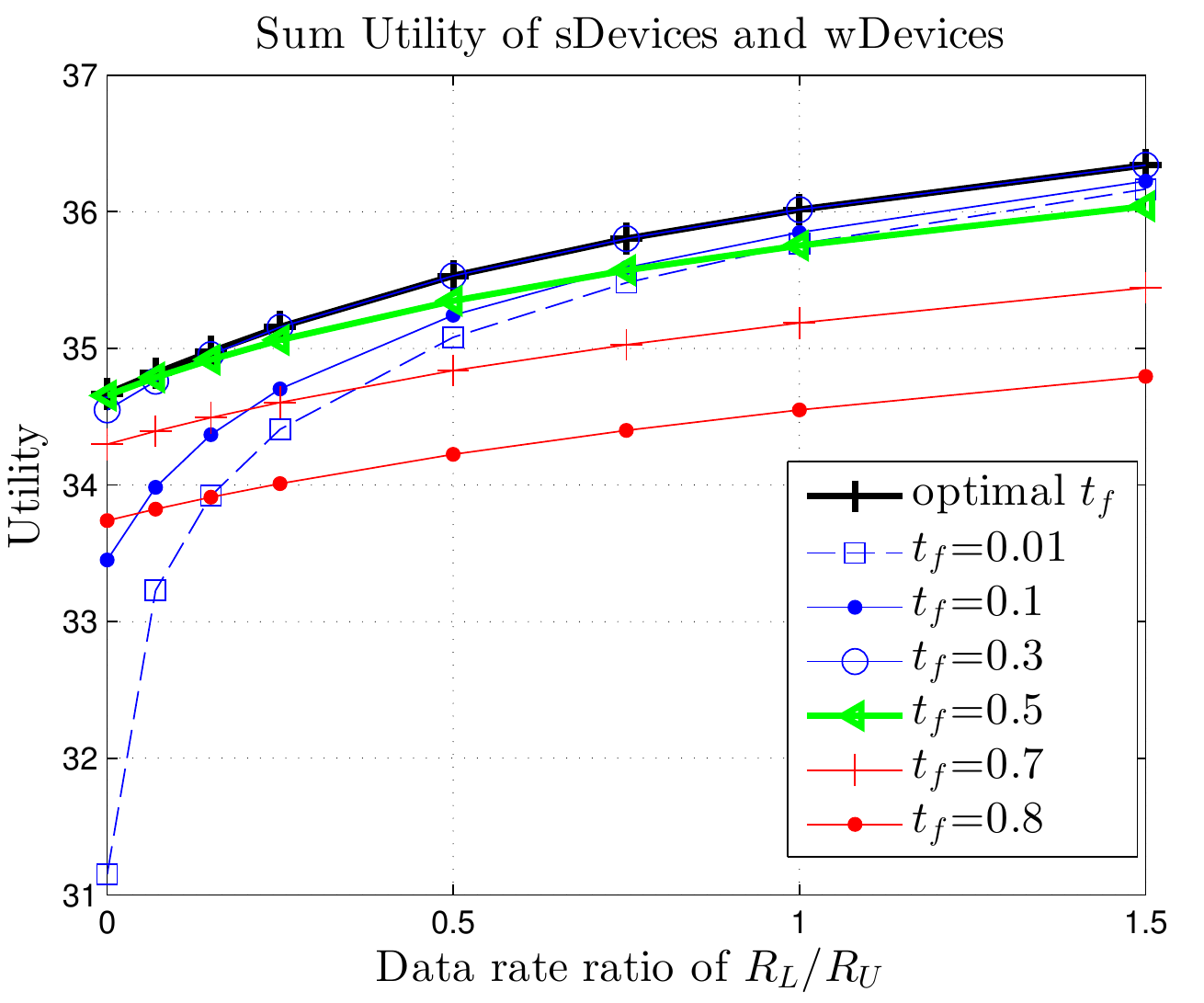}
        \label{fig:plot_utilities}
        }
\caption{ Optimal small cell usage, $t_f$, and sum utility as functions of $R^{\emph{tot}}_L/R_U$. Different ratios of $R^{\emph{tot}}_L/R_U$ are obtained by varying licensed bandwidth (1.4, 3, 5, 10, 15, 20 and 30 MHz). $t_{max}=0.9$. Fig. \ref{fig:plot_utilities} further assumes $N_w=1$ and $\bar{t}_w = 0.6$. } \label{fig:matlab_numerical_results}
\end{figure*}

\subsection{Intuitions Behind the Optimal Solution} \label{subsec:intuitions}
In this subsection, we discuss the intuitions behind the optimal $t_f$ solution.

\textbf{Case H1}: Firstly, we consider a high wDevice load case where
\begin{equation}
\bar{t}_w \geq t_{max}  \label{eq:cond1-intuition-caseA1}
\end{equation}
and
\begin{equation}
N_W
R^{\emph{tot}}_L /{R_U} \leq t_{max} \label{eq:cond2-intuition-caseA1}
\end{equation}
are both satisfied.
From (\ref{opt_prob:tf_expr})  and (\ref{opt_prob:tw_expr}) we have
\begin{eqnarray}\nonumber
t_f^* &=&  \frac{1}{N_W +1}  \left( t_{max} - N_W
\frac{R^{\emph{tot}}_L}{R_U}  \right),
\label{opt_prob:tf_expr-intuition1} \\
t_w^*&=&  \frac{N_W}{N_W +1}  (t_{max} + R^{\emph{tot}}_L / R_U) .  \label{opt_prob:tw_expr-intuition1}
\end{eqnarray}
The sDevice throughput is from both the licensed and unlicensed band. We can translate the total sDevice throughput into the unlicensed band channel time $t'_f$ as if the licensed band throughput were also obtained from the unlicensed band,
\begin{equation}
t'_f =\frac{R^{\emph{tot}}_L + t_f^* R_U}{R_U} = \frac{t_{max} + R^{\emph{tot}}_L / R_U}{N_W + 1}.
\end{equation}
As we can see,
\begin{equation}
t'_f = \frac{t'_f+t_w^*}{ N_W + 1}.
\end{equation}
Therefore, when the conditions (\ref{eq:cond1-intuition-caseA1}) and (\ref{eq:cond2-intuition-caseA1}) are both satisfied, the optimization process \emph{effectively translates the total sDevice throughput (of licensed and unlicensed bands) into the unlicensed band channel time} $t'_f$, and guarantees  that $t'_f$ is an \emph{equal share of the combined channel time} $(t'_f+t_w^*)$.
An extreme case of the condition (\ref{eq:cond2-intuition-caseA1})  is
\begin{equation}
N_W
R^{\emph{tot}}_L /{R_U} \ll t_{max}, \label{eq:cond3-intuition-caseA1}
\end{equation}
then we can obtain
\begin{eqnarray}
t_f^* & \approx &  \frac{t_{max}}{N_W +1}  ,
\label{opt_prob:tf_expr-intuition2} \\
t_w^*& \approx &  \frac{N_W t_{max}}{N_W +1}.  \label{opt_prob:tw_expr-intuition2}
\end{eqnarray}
From  (\ref{opt_prob:tf_expr-intuition2}) and (\ref{opt_prob:tw_expr-intuition2}), we further confirm that the optimization process assigns an equal share of the unlicensed band channel time to the sDevice.

\textbf{Case H2:} If  (\ref{eq:cond1-intuition-caseA1}) is satisfied (i.e., high wDevice load) and (\ref{eq:cond2-intuition-caseA1}) is not, we can get  $t^*_f = 0$ from (\ref{opt_prob:tf_expr}). Similar to Case H1, the optimization process in this case still translates the total sDevice throughput into the unlicensed band channel time $t'_f$, but $R^{\emph{tot}}_L /{R_U}$ is so large that we can never achieve  $t'_f = (t'_f+t_w^*)/(N_W +1)$. The best solution is $t^*_f = 0$ which minimizes the difference between $t'_f$ and $(t'_f+t_w^*)/(N_W +1)$.

\textbf{Case L1:} We consider a low wDevice load case where  the following is satisfied
\begin{equation}
\bar{t}_w \leq N_W \frac{{t_{max} + R^{\emph{tot}}_L /R_U}}{N_W +1}.  \label{eq:cond1-intuition-caseL1}
\end{equation}
The optimal solution $t_f$ in (\ref{opt_prob:tf_expr}) can be simplified to,
\begin{equation}
t^*_f = (t_{max} - \bar{t}_w)^+.  \label{eq:result-intuition-caseL1}
\end{equation}
In this case since the wDevice aggregate traffic load is limited, the sDevice tries to use the remaining available channel time.

\subsection{Numerical Results} \label{subsec:numerical-results}

Fig. \ref{fig:plot_tf} shows the numerical results of Eq. (\ref{opt_prob:tf_expr}) under different ratios of $R^{\emph{tot}}_L/R_U$.
In Figures \ref{fig:plot_tf} and \ref{fig:plot_utilities}, we assume that $P_f^{*(k)}$ and $\gamma_f^{(k)}$ values are such that sDevices have fixed spectral efficiency 3.9 bits/second/Hz for licensed and unlicensed bands, hence only the licensed and unlicensed bandwidth affects $R^{\emph{tot}}_L$ and $R_U$, respectively. We fix unlicensed  bandwidth to 20MHz, and vary licensed bandwidth (1.4, 3, 5, 10, 15, 20 and 30 MHz) to obtain different instantaneous licensed band rate $R^{\emph{tot}}_L$'s. The LTE licensed carrier frequency bandwidth can only be up to 20MHz; the 30MHz is due to carrier aggregation of multiple licensed carrier frequencies. The highest rate 72Mbps is used for the wDevice physical layer data rate $R_W$.

Fig. \ref{fig:plot_utilities} shows the numerical results of sum utility in Eq. (\ref{opt_prob:obj}) with respect to $t_f$. Here we assume $N_w=1$ and $\bar{t}_w = 0.6$.
From Fig. \ref{fig:plot_tf} we see that the optimal $t_f$ depends on many parameters. In this figure, we observe that a constant $t_f$ cannot achieve good sum utility under every condition.
Note that constant $t_f=0.3$ achieves almost the optimal sum utility, which is because it is close to the optimal $t_f$ range [0.3, 0.45] (see Fig. \ref{fig:plot_tf}). Since we use fixed spectral efficiency for the licensed band, the utility gain in the figure is only from $t_f$ optimization, not power control.
Therefore, while the existing studies \cite{femto-PowerControl-2010Sundeep}  show that licensed band power control is very useful for small cells, this figure suggests that when power control has been done for the licensed band, we can further improve user utility by time-sharing control of $t_f$ in the unlicensed band.
In addition, we also observe that the utility increases as $t_f$ becomes closer to the optimal value. We study the impact of the sum utility sensitivity to the change of $t_f$ in Section \ref{subsec:Sensitivity-Analysis}.

\subsection{Sensitivity Analysis} \label{subsec:Sensitivity-Analysis}
In this subsection, we analyze the impact of $t_f$  to the sum utility $U_{sum}$  formulated in (\ref{opt_prob:obj}). Unlike the optimization analysis in the previous subsection, the $t_f$ assignment in this subsection may not be optimal, hence we do not assume $t_f + t_w = t_{max}$ here. We consider two cases based on the $t_f$ value.

\textbf{Case A:} If condition (\ref{opt_prob:sensitivity-caseA-cond}) is satisfied,
\begin{equation}
t_{max} - t_f \leq  \bar{t}_w~ , 
 \label{opt_prob:sensitivity-caseA-cond}
\end{equation}
we have
\begin{equation}
t_w = t_{max} - t_f.
 \label{opt_prob:sensitivity-caseA-tw}
\end{equation}
Plugging (\ref{opt_prob:sensitivity-caseA-tw}) into the sum utility (\ref{opt_prob:obj})  and taking the derivative of $U_{sum}$ with respect to $t_f$, we obtain the first order approximation to the sum utility change $\Delta U_{sum}$ when $t_f$ is increased to $(t_f+\Delta t_f)$,

\begin{equation}
\Delta U_{sum} = \left (\frac{1}{(\sum_k R_L(P_f^{(k)} \gamma_f^{(k)}))/R_U + \tilde{t}_f} - \frac{N_W}{ t_{max} - \tilde{t}_f} \right ) \Delta t_f.
 \label{opt_prob:sensitivity-caseA-result}
\end{equation}
Here we use the middle point $\tilde{t}_f = t_f + \Delta t_f/2$ to improve the approximation.

\textbf{Case B:} If condition (\ref{opt_prob:sensitivity-caseA-cond}) is not satisfied, we have
\begin{equation}
t_w = \bar{t}_w.
 \label{opt_prob:sensitivity-caseB-tw}
\end{equation}
Submit (\ref{opt_prob:sensitivity-caseB-tw}) into the sum utility (\ref{opt_prob:obj})  and take the derivative of $U_{sum}$ with respect to $t_f$, we obtain
\begin{equation}
\Delta U_{sum} = \left (\frac{1}{(\sum_k R_L(P_f^{(k)} \gamma_f^{(k)}))/R_U + \tilde{t}_f} \right ) \Delta t_f,
 \label{opt_prob:sensitivity-caseB-result}
\end{equation}
where  $\tilde{t}_f= t_f + \Delta t_f/2$.

\begin{figure}
  \center
  \includegraphics[width=3.1in]{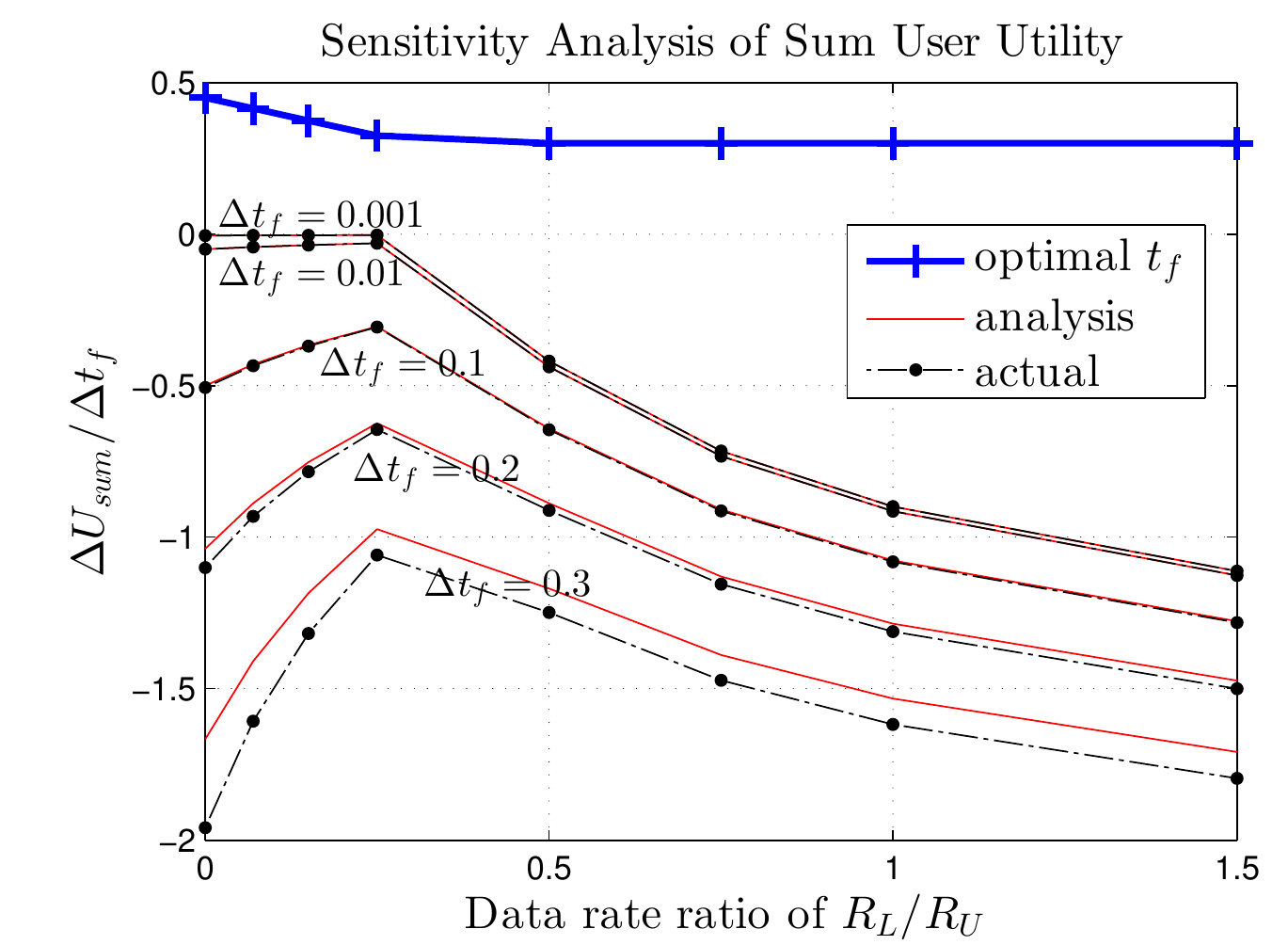}
  \caption{ Sensitivity of sum utility to variations in $t_f$ based on $N_w=1$, $\bar{t}_w = 0.6$ and optimal $t^*_f$. Note that the Y-axis is $\Delta U_{sum}/\Delta t_f$.  The ``analysis" curves are obtained from (\ref{opt_prob:sensitivity-caseA-result}) and (\ref{opt_prob:sensitivity-caseB-result}).   The ``actual" curves are obtained from (\ref{opt_prob:obj}) by calculating $(U_{sum}(t^*_f+\Delta t_f)-U_{sum}(t^*_f))/\Delta t_f$.  } \label{fig:sensitivity}
\end{figure}

Fig. \ref{fig:sensitivity} shows the numeric results of the above analysis. The ``actual" curves are obtained from (\ref{opt_prob:obj}) by calculating $(U_{sum}(t^*_f+\Delta t_f)-U_{sum}(t^*_f))/\Delta t_f$. The ``analysis" curves are obtained from (\ref{opt_prob:sensitivity-caseA-result}) and (\ref{opt_prob:sensitivity-caseB-result}). We observe that the sum utility degrades as $t_f$ deviates more from the optimal value, which is consistent with the observation in Fig. \ref{fig:plot_utilities}.
For negative $\Delta t_f$ values, we observe similar trend of sum utility degradation, which is not shown here.

\subsection{Implementation Issues} \label{subsec:Traffic-assign-Implementation}

The analysis in Section \ref{sec:opt_solution} considered  Shannon capacity as the rate function $R_L(\cdot)$. For practical LTE networks, Mogensen et al. \cite{LteRateApprox} propose a closed-form approximation for the LTE rate function,
\begin{eqnarray}
R_L(\textmd{SINR}) \approx \kappa_{bw} \cdot \kappa_{c} \cdot B
\log_2(1+\textmd{SINR} / \kappa_{sinr} ) ~~ \textmd{bits/s},
\label{implement:RL_approx}
\end{eqnarray}
where $\kappa_{bw}$ is the system efficiency that accounts for various system-level overheads including cyclic prefix,
pilot assisted channel estimation, and non-fully utilized frequency bandwidth (to prevent signal leakage to adjacent frequencies). The parameters $\kappa_{c}$ and $\kappa_{sinr}$ jointly adjust for the SINR implementation efficiency of LTE MCSs and receiver algorithms (e.g., linear, non-linear) \cite{LteRateApprox}. The value of $\kappa_{bw}$ can be directly derived from LTE protocol parameters, whereas $\kappa_{c}$ and $\kappa_{sinr}$ can be obtained by fitting the LTE rate curve generated from link-level simulations. For a realistic implementation of the traffic balancing strategy, the analysis in Section \ref{sec:opt_solution} can be replicated by using the $R_L(\cdot)$  in (\ref{implement:RL_approx}). In our simulations, we will approximate SISO LTE-A rates using $\kappa_{bw}=0.6726, \kappa_{c}=0.75$ and $\kappa_{sinr} = 1$ in (\ref{implement:RL_approx}).

Another implementation issue is that the proposed algorithm requires
the fBS to know the channel gain $|h_{fm}^{(k)}|$ of the fBS-to-mDevice
interference link, as required by the constraint
(\ref{opt_prob:cond_Imax}). In current cellular networks, exact
value of $|h_{fm}^{(k)}|$ may be difficult to obtain. However, this
problem will be addressed in the future, based on the current
developments in LTE-A. For example, the 3GPP is investigating
macro-Femto coordination mechanisms \cite{3GPP_TR_36.921_femto}.
Based on the mechanism proposed in \cite[Sec.\
7.2.2.6.2]{3GPP_TR_36.921_femto}, one can further estimate
$|h_{fm}^{(k)}|$ based on path loss. For beyond-4G cellular systems,
\cite{mBS_broadcastLocRB} proposes that the mBS broadcasts mDevice
locations and the resources used by each mDevice, so that
femtocells/small cells can estimate the $|h_{fm}^{(k)}|$ according
to the fBS-to-mDevice distances. In our simulations, we will assume
$|h_{fm}^{(k)}|$ is known at the fBS.

The fBS also needs to know the traffic load $\bar{t}_w $ of the
wDevices, which is the aggregate DL and UL channel time used by the wDevices before
the sDevice uses the unlicensed band, and can be obtained via fBS
long-term sensing.\footnote{Since the IFW fBS is connected to both sDevice and wDevices and the wDevices have both downlink and uplink traffic, the fBS transmission is split between the sDevice and wDevices. Therefore, the $\bar{t}_w $  measurement in IFW cell should take into account the DL transmission from the fBS to the wDevices and the UL transmission from the wDevices.} In addition, the fBS also needs to learn $N_W$,
the number of coexisting wDevices.  The IFW fBS can learn $N_W$
from the MAC addresses of the wDevices; whereas the DBF fBS can
learn $N_W$  using the RF fingerprint technique
\cite{RF-fingerprint} which allows the fBS to distinguish each
transmitter based on their unique radio characteristics (or ``RF
fingerprint") without decoding their signals.

\section{Adjusting Unlicensed Band Usage for sDevice in Integrated  Femto-WiFi}\label{sec_IFW}

The optimal traffic-balancing scheme introduced in Section \ref{sec:DL-assignment} requires the fBS to adjust $t_f$, the fraction of time the fBS transmits in the unlicensed band to the sDevice. In this section, we discuss how this can be done for the  IFW use case; in Section \ref{sec_DBF}, we will study the DBF use case.

Recall that in the IFW use case, the fBS contends with $N_W$ wDevices for the same channel. In addition, the fBS transmits to both sDevice and wDevices. The total fBS channel time is
\begin{equation}
t_{\emph{fBS}} = t_f + t_w^{dl} ,
 \label{eq:t_fBS}
\end{equation}
where  $t_w^{dl}$ is the DL channel time for all wDevices. When $t_w=\bar{t}_w$, $t_w^{dl}$ is the time required to transmit the DL traffic load of all wDevices.   When $t_w<\bar{t}_w$,  $t_w^{dl}$ is a predefined fraction of $t_w$; in this paper, we assume $t_w^{dl}$ is proportional to the DL traffic load $\bar{t}_w^{dl}$,
\begin{equation}
t_w^{dl} = \frac{\bar{t}_w^{dl}}{\bar{t}_w} t_w.
 \label{eq:t_wdl}
\end{equation}
Once the fBS obtains $t_{\emph{fBS}}$ channel time, it assigns $t_f$ to the sDevice and $t_w^{dl}$ to the wDevices.

Studies such as \cite{solve_WiFi_Anomaly_letter_AdjustW} have shown that
WiFi channel usage in the unlicensed band
is a monotonically decreasing function of its initial
backoff window size.
Therefore, we can adjust the fBS channel usage $t_{\emph{fBS}}$ in the unlicensed band by tuning the fBS initial backoff window size $W_f$.
However, directly using the analytical result \cite{solve_WiFi_Anomaly_letter_AdjustW} in practice
 requires the knowledge of the initial
backoff window sizes and the transmission durations of the wDevices, which are possible via the feedback from the wDevice to the IFW fBS. However, such feedback is not supported by existing WiFi protocol. Therefore, to ensure that our algorithm works for existing WiFi technologies, we
developed the bisection-based Algorithm \ref{algorithm:IFW-Find-Wf}
which only uses \emph{the trend that the fBS channel usage $t_{\emph{fBS}}$ in
the unlicensed band is a monotonic function of its initial backoff
window size $W_f$.} The monotonicity of the $t_{\emph{fBS}}(W_f)$ function
makes it appropriate to use bisection
\cite{Computing-textbook-bisection} to efficiently search for the
$W_f$ that can achieve a given $t_{\emph{fBS}}$. The bisection algorithm needs
to know the channel usages for some window size values, which can be
obtained through measurements. The fBS first sets its parameter
$W_f$ to the window size value that needs to be measured, and
utilizes the channel for a certain time. During this time, the fBS
periodically records its transmission state (idle or transmitting)
at each sampled time instant. We denote $N_{tot}$ as the total
number of time samples, out of which $N_{tx}$ samples turn out that
the fBS is transmitting. Then the measured $t_{\emph{fBS}}$ is
\begin{equation}
t_{\emph{fBS}} = N_{tx} / N_{tot} ,  \label{eq:IFW-how-to-measure-tf}
\end{equation}
The measurements can be implemented in software and do not require
any additional hardware. Algorithm \ref{algorithm:IFW-Find-Wf} is
guaranteed to converge to the desired $W_f^*$ and $t_{\emph{fBS}}^*$ by
\cite{Computing-textbook-bisection}.

\algsetup{indent=2em}
\begin{algorithm}
\caption{Find the desired $W_f^*$ for IFW to obtain channel usage
$t^*_{\emph{fBS}}=t_f^* +t_w^{dl}$}\label{algorithm:IFW-Find-Wf}
\begin{algorithmic}[1]
%


\STATE Initialize $W_f$ range $[W_f^{(1)}, W_f^{(2)}]$, such that
$t^*_{\emph{fBS}} \in [t_{\emph{fBS}}(W_f^{(2)}),t_{\emph{fBS}}(W_f^{(1)})]$. \REPEAT
    \STATE Set $W_f = (W_f^{(1)} + W_f^{(2)})/2$ and measure $t_{\emph{fBS}}$.
    \IF {$(t_{\emph{fBS}}(W_f) - t_{\emph{fBS}}^*) (t_{\emph{fBS}}(W_f^{(1)}) - t_{\emph{fBS}}^*) > 0$}
        \STATE Set $W_f^{(1)} = W_f$.
    \ELSE
         \STATE Set $W_f^{(2)} = W_f$.
    \ENDIF
\UNTIL{$|t_{\emph{fBS}}(W_f) - t_{\emph{fBS}}^*| <$ tolerance}

\end{algorithmic}
\end{algorithm}

\section{Adjusting Unlicensed Band Usage for Dual-Band Femtocell}\label{sec_DBF}
In this section, we discuss how $t_f$ can be adjusted for the DBF which uses the access scheme introduced in Section \ref{sec: ch-access-scheme}.

\subsection{DBF Channel Usage Analysis}\label{subsection-DBF-ch-usage-analysis}

We first describe an analytical model to obtain $t_f$ in DBF.
In the unlicensed band, the DBF fBS contends with $N_W$ wDevices and one WiFi AP for the channel. Recall that we denote the network formed by the wDevices and the AP as ``WLAN".
According to the channel access scheme in Section \ref{sec: ch-access-scheme}, in a channel access attempt, if the channel is successfully obtained, the
fBS transmits for a fixed duration $T_{\textmd{cellTx}}$; otherwise,
it will attempt again after a fixed time duration
$T_{\textmd{attempt}}$. The success probability for each attempt is denoted as $P_{\textmd{DBFsuc}}$. Recall that the fBS channel access attempts are
statistically independent. As such, $1/P_{\textmd{DBFsuc}}$
attempts are needed on average for the fBS to obtain the channel. Considering $T_\textmd{sensing} \ll \min(T_{\textmd{cellTx}},  T_{\textmd{attempt}})$, the fraction
of channel time occupied by the small cell is
\begin{equation}
t_f =    \frac{T_{\textmd{cellTx}}} {(1/P_{\textmd{DBFsuc}}) \cdot
T_{\textmd{attempt}} + T_{\textmd{cellTx}} }
 =    \frac{\eta}{1/P_{\textmd{DBFsuc}} +  \eta  },
\label{eq:DBF-tf-expression}
\end{equation}
where
\begin{equation}
\eta  =    T_{\textmd{cellTx}}  /  T_{\textmd{attempt}}.
\label{eq:DBF-eta-protocol-def} \nonumber
\end{equation}
As we can see from (\ref{eq:DBF-tf-expression}), to find $t_f$, we need to know the attempt success probability $P_{\textmd{DBFsuc}}$ which will be obtained in the following.

WiFi nodes (wDevices or AP) access the channel in a random fashion, resulting in
random channel states (idle, collision, or successful transmission)
at any given time. The fractions of time that WiFi channel is idle,
in collision state and in successful transmission state, are mainly
determined by transmission buffer status, number of contenders and
exponential backoff parameters. Recall that under the proposed
traffic-balancing scheme in Sec. \ref{sec:DL-assignment},  the wDevices
will always have data to send, although its traffic load may be
limited. In addition, since there are many wDevices, the
introduction of a fBS increases the number of contenders by a small
fraction. Hence, the WLAN has almost the same fractions of idle,
collision and successful transmission time, respectively, in the
WLAN/DBF cell coexistence scenario as in the full-buffer WLAN-only
scenario.

The fraction of idle channel time in a full-buffer WLAN-only network
can be obtained by using a 2D Markov chain to analyze the WLAN
exponential backoff process, as done in
\cite{Bianchi,WiFi_modeling_Foh_freezingCounter,WiFi_analysis_freezingCounter_Lee2005Conf}.
Foh and Tantra's analysis \cite{WiFi_modeling_Foh_freezingCounter}
show that the probabilities for a WLAN channel to be in idle,
collision and successful transmission states are related to the
previous channel state. Given that the previous channel state is
busy (i.e., successful transmission or collision), the conditional
probabilities for these three states are $P_I$, $P_c$ and $P_s$,
respectively. Given that the previous channel state is  idle, the
conditional probabilities for these three states are $Q_I$, $Q_c$
and $Q_s$, respectively. These probabilities can be obtained using
the analytical results in \cite{WiFi_modeling_Foh_freezingCounter}
(see equations (2)-(5) in \cite{WiFi_modeling_Foh_freezingCounter}).


\begin{figure*}
  \center
  \includegraphics[width=6.0in]{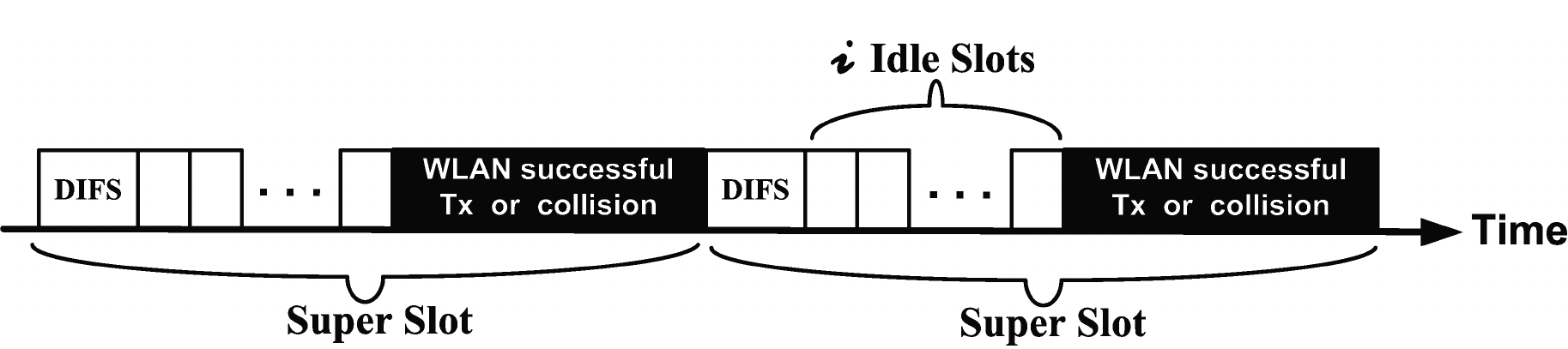}
  \caption{WLAN channel can be viewed as consisting of consecutive super slots.} \label{fig_superslot}
\end{figure*}

When the channel is busy, every WLAN node (wDevice or AP) freezes
its backoff counter and waits until the channel becomes idle. Then
each node will defer for a fixed duration named DCF Interframe
Space (DIFS) before resuming counting down its backoff counter in
every idle backoff slot.  The nodes that hit zero will transmit and
the other ones will freeze their counters. A transmission may be
successful or collided. As illustrated by Fig. \ref{fig_superslot},
we denote a \emph{super slot} (SS) as a WLAN time period consisting
of a DIFS,  $i$ ($i = 0, 1, 2, \dots$) consecutive idle backoff time
slots, and a busy channel state. The WLAN channel is constituted by
consecutive SS's, so an fBS attempt time instant is randomly located
in one of the SS's. We denote $SS^{(s)}_{i,u}$ as an SS with $i$
consecutive idle slots and a successful transmission from or to device
$u$. We also denote $SS^{(c)}_{i}$ as an SS with $i$ consecutive idle
slots and a collided transmission. Since each WLAN node has an equal
number of transmissions \cite{equal_access_measure_infocom2003},
given that the previous channel state is busy, the probability that
we have a successful transmission from device $u$ is $P_s/N_W$.
Similarly, given that the previous channel state is idle, the
probability that we have a successful transmission from device $u$ is
$Q_s/N_W$. Hence, the probability of observing $SS^{(s)}_{i,u}$ is
\begin{eqnarray}
P^{(s)}_{i,u} =  \left\{ \begin{array}{l}
P_s / N_W, ~~~~~~~~~~~ \textmd{if}~ i=0; ~~ \textmd{and} \\
P_I Q_I^{i-1}Q_s / N_W, ~ \textmd{if}~ i=1, 2, \dots .
\end{array} \right .
\end{eqnarray}

Likewise, the probability for $SS^{(c)}_{i}$ to happen is
\begin{eqnarray}
P^{(c)}_{i} =  \left\{ \begin{array}{l}
P_c, ~~~~~~~~~~~~~ \textmd{if}~ i=0; ~ \textmd{and} \\
P_I Q_I^{i-1}Q_c, ~~~ \textmd{if}~ i=1, 2, \dots .
\end{array} \right .
\end{eqnarray}

Further denoting the durations of a DIFS, an idle backoff slot,  a
collision and a successful transmission from device $u$ ($u=1, 2,
\dots, N_W$), as $T_d$, $T_I$, $T_c$ and $T_{s,u}$, respectively,
the durations of $SS^{(s)}_{i,u}$ and $SS^{(c)}_{i}$ are
\begin{eqnarray}
T^{(s)}_{i,u} &=& T_d + i T_I + T_{s,u} ~~ \textmd{and}
\label{eq:T-SuperSlot-c} \\
T^{(c)}_{i} &=& T_d + i T_I + T_c  ,  ~~ \textmd{respectively.}
\label{eq:T-SuperSlot-c}
\end{eqnarray}
Here $T_{s,u}$ is a constant that mainly depends on the payload
length and data rate of device $u$, as well as various protocol
overheads such as RTS/CTS/ACK messages. The constants $T_d$ and
$T_I$ are defined in IEEE 802.11 standards. The expected duration of
an SS is
\begin{eqnarray}
T_{\textmd{avg}} &=& \sum_{i=0}^{\infty} P^{(c)}_{i} T^{(c)}_{i} +   \sum_{i=0}^{\infty} \sum_{u=1}^{N_W} P^{(s)}_{i,u} T^{(s)}_{i,u} \nonumber \\
&=&   T_d + \frac{ P_I T_I }{1- Q_I} +   T_c \left( P_c + \frac{Q_c
P_I}{1- Q_I} \right )   +  \nonumber \\
&& \frac{ \sum_u T_{s,u} }{N_W} \left( P_s + \frac{Q_s P_I}{1- Q_I}
\right )    . \label{eq:T_AvgSuperSlot}
\end{eqnarray}
Therefore, the probability that an fBS attempt time is located in an
$SS^{(c)}_{i}$ and $SS^{(s)}_{i,u}$ are respectively
\begin{eqnarray}
\tilde{P}^{(c)}_{i} &=&
 \frac{ P^{(c)}_{i}T^{(c)}_{i}  }{ T_{\textmd{avg}} }
  ~~~~~  \textmd{and}  ~~~~~
\tilde{P}^{(s)}_{(i,u)}
 = \frac{ P^{(s)}_{(i,u)} T^{(s)}_{i,u}  }{ T_{\textmd{avg}} } .
\end{eqnarray}

We denote $i_0$ as the minimum number of idle slots that a $SS^{(s)}_{i,u}$ must have in order to provide a long enough idle period so that an fBS channel sensing may be successful,
\begin{equation}
   i_0 = \left\lceil \frac{T_{\textmd{sensing}}-T_d}{T_I} \right\rceil^+.
\end{equation}
Here $\lceil x \rceil^+$ denotes the smallest non-negative integer that is
greater than or equal to $x$.
Given that an fBS attempt time instant is located in an $SS^{(c)}_{i}$ and $i \geq i_0 $, the conditional probability that the fBS attempt is successful is
\begin{eqnarray}
P^{(c)}_{\textmd{Suc},i} &=&  \frac{T_d + i T_I  - T_{\textmd{sensing}}}{T^{(c)}_{i}} ,
\end{eqnarray}
which is also the conditional probability that the channel is idle for at least $T_{\textmd{sensing}}$ time after a DBF attempt time instant.
Likewise, given that an fBS attempt is in an $SS^{(s)}_{i,u}$ and $i \geq i_0 $, the conditional fBS attempt success probability is
\begin{eqnarray}
P^{(s)}_{\textmd{Suc},(i,u)} &=&  \frac{T_d + i T_I  - T_{\textmd{sensing}}}{ T^{(s)}_{i,u} }  .
\end{eqnarray}

Then the probability that the DBF fBS successfully obtains the channel
is:
\begin{eqnarray}
&&P_{\textmd{DBFsuc}} =
\sum_{i=i_0}^{\infty} \tilde{P}^{(c)}_{i} P^{(c)}_{\textmd{Suc},i}  + \sum_{i=i_0}^{\infty} \sum_{u=1}^{N_W} \tilde{P}^{(s)}_{(i,u)} P^{(s)}_{\textmd{Suc},(i,u)}  \nonumber \\
&&=
 \left\{ \begin{array}{l}
P_I Q_I^{i_0 -1}(T_d + i_0 T_I  - T_{\textmd{sensing}} + \frac{T_I Q_I}{1- Q_I}) /  T_{\textmd{avg}}, \\
~~~~~~~~~~~~~~~~~~~~~~~~~~~~~~~~~~~~\textmd{if} ~ i_0 \geq 1  \\
(T_d - T_{\textmd{sensing}} + \frac{T_I P_I}{1- Q_I}) /
T_{\textmd{avg}}, ~~ \textmd{if} ~  i_0 =0 .
\end{array} \right . 
\label{eq:DBF-PDBFsuc-fully}
\end{eqnarray}

Note that $P_{\textmd{DBFsuc}}$ does not depend on
$T_{\textmd{attempt}}$ and $T_{\textmd{cellTx}}$. We can then obtain
$t_f$ from (\ref{eq:DBF-tf-expression}) using the
$P_{\textmd{DBFsuc}}$ computed in (\ref{eq:DBF-PDBFsuc-fully}).

The analysis above 
provides an exact relationship between $t_f$ and the parameters of
both DBF and the non-cellular WLAN.
In particular, it is shown in (\ref{eq:DBF-tf-expression}) that
$t_f$ depends on $\eta$ and $P_{\textmd{DBFsuc}}$, where $\eta$ is a
DBF channel access parameter. $P_{\textmd{DBFsuc}}$ as shown in
(\ref{eq:DBF-PDBFsuc-fully}) is a function of the non-cellular WLAN
parameters, $N_W$, $T_I$, $P_I$, $Q_I$ and $T_{\textmd{avg}}$, the
latter three of which can be computed when the data rates and
payload lengths of all the wDevices and the AP are known. While this
analytical relationship helps us understand how $t_f$ can be
adjusted by varying DBF channel access parameter $\eta$, to directly apply it
in practice, data rates and payload lengths of the AP and wDevices need
to be obtained.  As an alternative, we propose the following
practical method to obtain $P_{\textmd{DBFsuc}}$. The fBS can
directly estimate $P_{\textmd{DBFsuc}}$ based on its recent channel
access records,
\begin{equation}
P_{\textmd{DBFsuc}} = N_{s} / N_{\textmd{attempt}} ,
\label{eq:DBF-how-to-measure-PDBFsuc}
\end{equation}
where $N_{\textmd{attempt}}$ is the total number of channel access
attempts,  out of which $N_{s}$ turn out to be successful.  This
method requires no parameter knowledge of wDevices. Using the $P_{\textmd{DBFsuc}}$ obtained in
(\ref{eq:DBF-how-to-measure-PDBFsuc}), the fBS can further obtain
$t_f$ via (\ref{eq:DBF-tf-expression}). This practical method will
be used to obtain the traffic balancing simulation results reported
in Section \ref{sec:Femto-system-simulations}.

\subsection{Validation of the Analysis}\label{subsection-DBF-sim}

We carry out a simulation study to validate the analysis in Sec.
\ref{subsection-DBF-ch-usage-analysis}. We consider a WLAN
consisting of one AP and three wDevices, and a DBF consisting of one
fBS and one sDevice.  Recall that the DBF uses the unlicensed band for
downlink traffic only, whereas the WLAN uses the unlicensed band in
both uplink and downlink. The analytical and simulation results are
obtained based on a fixed WiFi data rate 72Mbps and a packet payload
length of 1500 bytes.

\begin{figure}
  \center
  \includegraphics[width=3.1in]{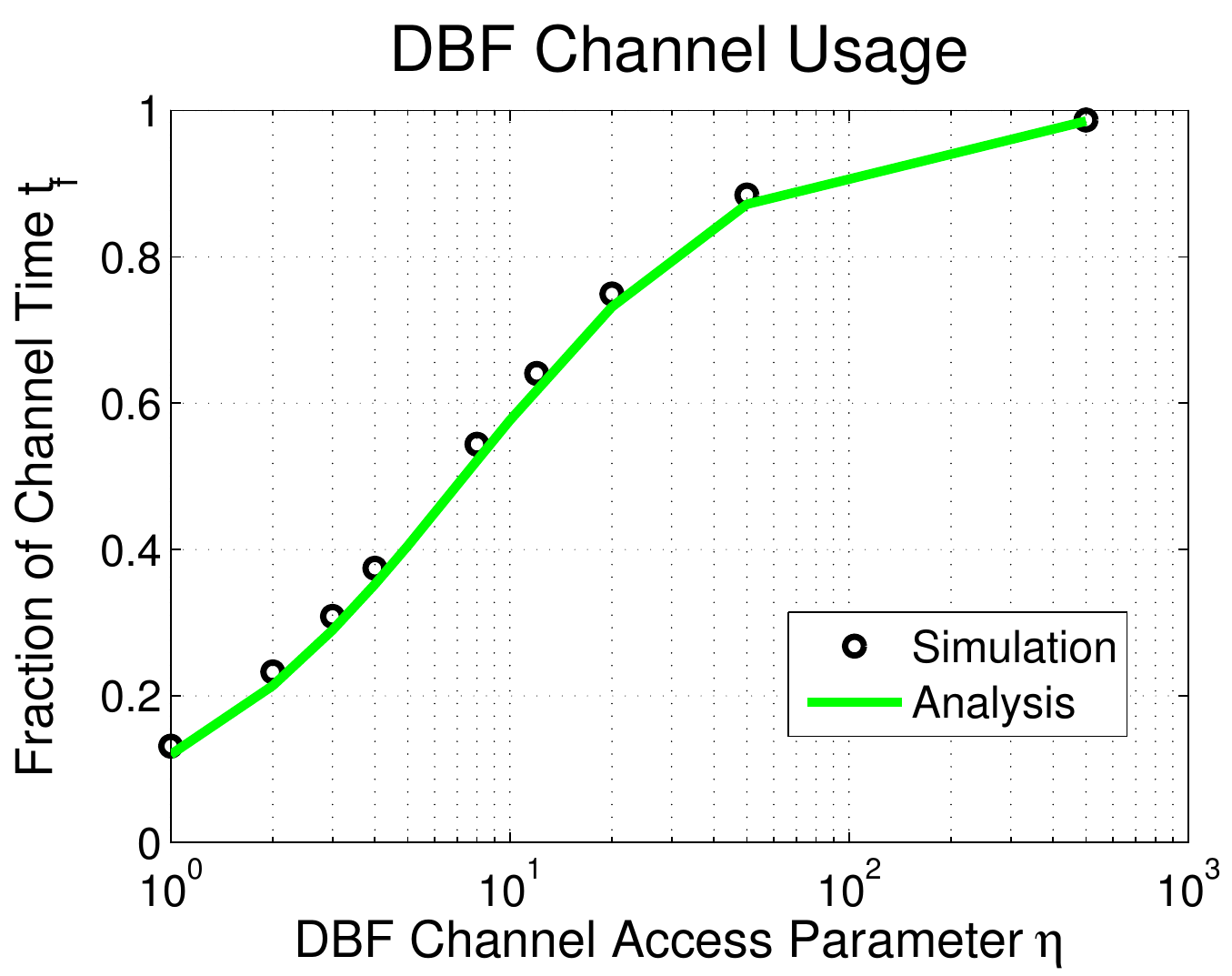}
  \caption{DBF channel usage $t_f$ from analysis and simulations.
  Markers are simulation results and curves are the numerical
  results of the analysis in this section (Sec. \ref{subsection-DBF-ch-usage-analysis}). }  \label{fig:DBF-sim}
\end{figure}

We fix $T_{\textmd{attempt}}$ to 1ms and vary $T_{\textmd{cellTx}}$
from 1ms to 500ms, hence $\eta$ varies from 1 to 500. The impact of
$\eta$ on DBF channel usage $t_f$ is shown in Fig.
\ref{fig:DBF-sim}. The curve labeled ``Analysis" is obtained from
(\ref{eq:DBF-tf-expression}) using the $P_{\textmd{DBFsuc}}$
computed in (\ref{eq:DBF-PDBFsuc-fully}). Recall that all parameters
in (\ref{eq:DBF-PDBFsuc-fully}) can be computed as functions of the
data rates and payloads of all WLAN devices and WLAN protocol
parameters provided in the 802.11 standards. The curve labeled
``Simulation" is obtained from the simulation data for $t_f$.

We observe from  Fig. \ref{fig:DBF-sim} that the simulation results
match analytical results well. As predicted by
(\ref{eq:DBF-tf-expression}), DBF channel time $t_f$ is an
increasing function of $\eta$.
Additional simulations with different $T_{\textmd{attempt}}$ values lead to
similar results as shown in Fig.
\ref{fig:DBF-sim} and are not shown here. This demonstrates that
$\eta$ is the main DBF parameter that impacts DBF channel usage, which is consistent with our analytical results
in (\ref{eq:DBF-tf-expression}).

\section{Performance Evaluation}\label{sec:Femto-system-simulations}

In this section, we evaluate the proposed traffic balancing strategy
in practical deployment scenarios. The simulation platform in
\cite{CoopMAC_WirelessMag2006}\cite{EqualAccess_CoopMAC_Jsac2007}, a
customized event-driven IEEE 802.11 network simulator built in C
language, is extended to simulate the activities and interactions of
macrocell,  small cell and non-cellular WLANs in both licensed and
unlicensed bands. The simulator models random packet arrivals at the
IP layer, channel access schemes (for macrocell, small cell and
WLAN) at the MAC layer, and includes interference computation and
SINR-to-throughput mappings at the PHY layer. Table
\ref{table:parameters} summarizes the path loss models and
parameters used in the simulations. The path loss models are based
on a Small Cell Forum whitepaper \cite{femto-forum-whitepaper}, where
path loss (PL) is in dB, distance $R$ is  in meters, and $L_{ow}$ is
the outer wall penetration loss.

\begin{table}
\caption{Parameters Used in Simulations} 
\centering
\begin{tabular}{|ll|}
\hline
 $t_{max}$ = 0.9  &   No. of Lic. subchannels $K$: 30   \\
\hline
\multicolumn{2}{|l|}{ $\bar{I}_k =$ -100 dBm,  $k = 1, 2, \dots, K$  } \\
\hline
IP Packet Size:   & 1500 Bytes \\
\hline
Transmit Power  &            \\
mBS:      40 dBm   &      fBS/WLAN AP:      15 dBm \\
\hline
Noise Power: &  -95 dBm (over 20MHz BW)          \\
\hline

\multicolumn{2}{|l|}{\textbf{Path Loss Models} for Licensed and Unlicensed Bands}      \\
\hline \multicolumn{2}{|l|}{mBS $\leftrightarrow$ mDevice }      \\
\multicolumn{2}{|l|}{PL = 15.3 + 37.6 log$_{10}(R)$ }      \\
\hline \multicolumn{2}{|l|}{mBS or mDevice $\leftrightarrow$ fBS or sDevice }      \\
\multicolumn{2}{|l|}{PL = 15.3 + 37.6 log$_{10}(R)$ + $L_{ow}$, $L_{ow}=10$dB  }     \\
\hline
\multicolumn{2}{|l|}{fBS $\leftrightarrow$ associated sDevices }     \\
\multicolumn{2}{|l|}{WLAN AP $\leftrightarrow$ associated stations  }      \\
\multicolumn{2}{|l|}{PL = 38.46 + 20 log$_{10}(R)$ + $0.7R$ }\\
\hline
\multicolumn{2}{|l|}{fBS or sDevice $\leftrightarrow$ fBS or sDevice in different cells, }      \\
\multicolumn{2}{|l|}{AP or station $\leftrightarrow$ AP or station in different WLANs, }   \\
\multicolumn{2}{|l|}{fBS or sDevice $\leftrightarrow$ WLAN AP or station  }    \\
\multicolumn{2}{|l|}{PL = 15.3 + 37.6 log$_{10}(R)$ + $L_{ow}$, $L_{ow}=20$dB  }     \\
\hline
\end{tabular}
\label{table:parameters}
\end{table}


LTE-Advanced \cite{LTE-A} is adopted as the cellular air interface while
802.11n \cite{802.11n} with a frame aggregation level of 15K Bytes is used for the WiFi air interface.
The bandwidth of WiFi is set 20 MHz.
The approximate LTE rate function described in Section \ref{subsec:Traffic-assign-Implementation} is used in the simulations.

We consider the following four use cases where a user or a group of users simultaneously use multiple sDevices and wDevices as shown in Fig. \ref{table:4UseCases}.
\begin{itemize}
\item   Case 1 (Cellular WiFi Hotspot): Each small cell operates
only in  the unlicensed band using 802.11n air interface. The WiFi hotspot is used by both sDevices and wDevices.


\item   Case 2 (Separate Femto+WLAN): Each small cell (femtocell) operates only
in licensed bands with the LTE air interface. In order to serve wDevices, the femtocell is deployed together with a WiFi AP. The femto BS and the WiFi AP are physically separate.

\item   Case 3 (IFW): Each small cell (IFW) \cite{femto-forum-IFW}
operates in licensed and unlicensed bands with LTE and WiFi air
interfaces, respectively. The IFW is used by both sDevices and wDevices; there is no need to deploy additional WiFi APs. The simplest scheme (``IFW, Simple") uses equal power in all subchannels and fixed $t_f$ (set to 0.8). We also consider the optimal traffic balancing strategy described in Section \ref{sec:DL-assignment} (``IFW, Optimal").

\item   Case 4 (DBF+WLAN): Each small cell (DBF)
operates in both licensed and unlicensed bands with LTE air
interface. In order to serve wDevices, the DBF is deployed together with a WiFi AP. The simplest scheme (``DBF+WLAN, Simple") uses equal power in all subchannels and fixed $t_f$ (set to 0.8). We also consider the optimal traffic balancing strategy described in Section \ref{sec:DL-assignment} (``DBF+WLAN, Optimal"). The DBF channel access scheme in the unlicensed band is described in Section \ref{sec: ch-access-scheme}.

\end{itemize}

\begin{figure}
  \center
  \includegraphics[width=3.1in]{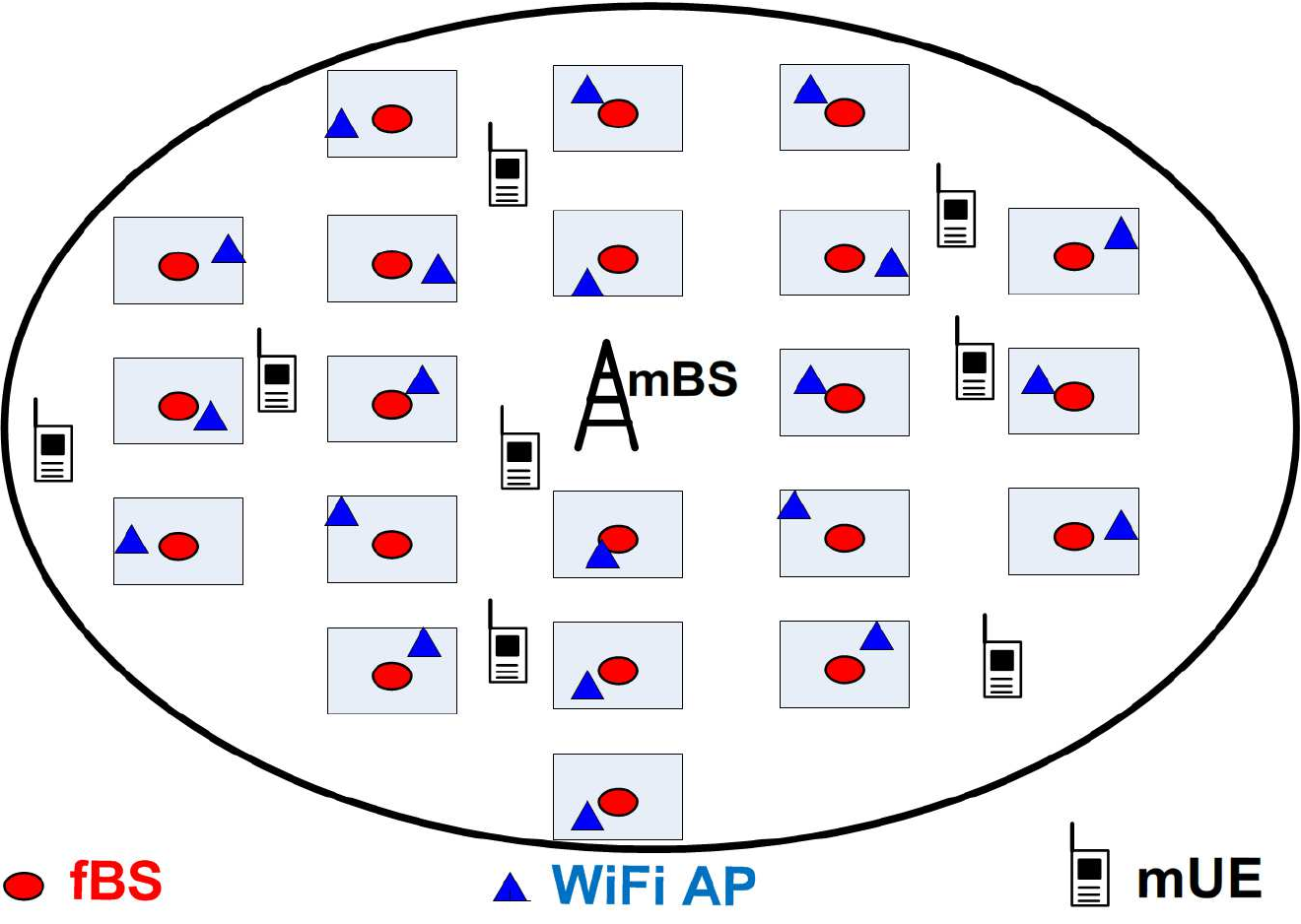}
  \caption{The network topology used in simulations (sDevices and wDevices are not shown). The WiFi AP is shown for Cases 2 and 4 only; there is no WiFi AP in Case 1 or 3.} \label{fig_topology}
\end{figure}

\subsection{A simple scenario}
We first consider a very simple scenario, in order to understand the intuitions behind our algorithms better.
As aforementioned, the existing studies \cite{femto-PowerControl-2010Sundeep} mainly focus on power control; whereas this work focuses on both time-sharing control (in the unlicensed bands) and power control (in the licensed band). To understand the gain from time-sharing control alone, we assume there is only one fBS, one sDevice, one wDevice and no macrocell, which eliminates inter-cell interference in the licensed band. Recall that we do not consider fading, so the optimal power allocation scheme is the same as the ``simple" scheme that assigns equal power to the subchannels. In addition, In Case 1, the short-range user only uses the unlicensed band; while in the other three cases, the user uses both licensed and unlicensed bands. To minimize the impact of the small cell frequency bandwidth imbalance between Case 1 and the other three cases, we assume that LTE licensed bandwidth is 1.4MHz which is the lowest allowed LTE bandwidth. Note that the unlicensed bandwidth is 20MHz. The sDevice and wDevice downlink traffic loads are 300Mbps (always full buffer) and 35Mbps (may not always be full buffer), respectively. Uplink traffic of each device is disabled to simplify the scenario.

Table (\ref{table:sim_1fBS1AP-noMacro}) shows the simulation results, from which we have the following observations.
\begin{itemize}
  \item The ``Separate Femto+WLAN" case has much lower sum throughput and sum utility than the other cases, because the sDevice can only use the 1.4MHz licensed band, whereas the sDevice in the other cases can share the 20MHz unlicensed band with the wDevice.
  \item DBF cases have higher sum throughput than the other cases, because of the higher efficiency of LTE than WiFi at the MAC layer. For 20MHz bandwidth, the simulator assumes that LTE and WiFi physical layer rates are 78Mbps and 72Mbps, respectively. But the maximum achievable MAC layer throughput are 75Mbps and 61Mbps, respectively. This is mainly because WiFi MAC protocol is distributed and contention-based, which incurs much channel access overhead (e.g., random backoff); whereas LTE MAC is centralized where the network schedules resources for each device.
  \item Compared with ``DBF+WLAN, Simple" case, ``DBF+WLAN, Optimal" case has lower sum throughput but higher sum utility. This is mainly due to the traffic balancing in the unlicensed band: $t_f=0.8$ is used by ``DBF+WLAN, Simple" case, whereas in the ``DBF+WLAN, Optimal" case, the intuitions from Section \ref{subsec:intuitions} suggest that our traffic balancing algorithm translates the total sDevice throughput (of licensed and unlicensed bands) into the unlicensed band channel time $t'_f$ (=38Mbps/78Mbps=0.49), and tries to guarantee  that $t'_f$ is an \emph{equal share of the combined channel time} $(t'_f+t_w^*)$ ($t_w^*$=33.7Mbps/72Mbps=0.47). The resulting optimal $t^*_f$ is 0.42 in this case.
       The same observation can be made by comparing results between``IFW, Simple" and ``IFW, Optimal" cases.
  \item In ``WiFi hotspot" case, the throughput disparity between sDevice and wDevice is small, due to the natural long-term fairness of WiFi MAC protocol (sDevice throughput is slightly higher than wDevice due to higher sDevice traffic load), so the effective $t_f$ is close to the optimal $t^*_f$ (=0.45) for zero-Hz licensed bandwidth  (see Fig. \ref{fig:plot_tf}). As a result, the sum utility (34.5) is very close to that of the ``IFW, Optimal" case (34.6).

\end{itemize}

\begin{table*}[ht]
\caption{Simulations results for scenario 1:  No macrocell (only one fBS, one WiFi AP, one sDevice and one wDevice), LTE Licensed Bandwidth 1.4MHz, wDevice Load 35Mbps} 
\centering
\begin{tabular}{|l|c|c|c|c|c|c|}
\hline
\textbf{Use} &WiFi   &   Separate   &  IFW   &  IFW   &  DBF+WLAN   & DBF+WLAN    \\
\textbf{Cases} & Hotspot  &    Femto+WLAN  &   Simple  &   Optimal  &   Simple  &  Optimal   \\\hline
sDevice &	 & 	 & 	 & 	 & 	 &    \\
Throughput  &32.8	 & 5.5	 & 51.7	 & 30.7	 & 66.9	 & 38.0   \\
(Mbps) &	 & 	 & 	 & 	 & 	 &    \\
\hline
wDevice &	 & 	 & 	 & 	 & 	 &    \\
Throughput & 28.5&	35.0&	11.6&	35.0&	11.7&	33.7   \\
(Mbps) &	 & 	 & 	 & 	 & 	 &    \\
\hline
Sum device &	 & 	 & 	 & 	 & 	 &    \\
Throughput &61.3 & 	40.5 & 	63.3 & 	65.7 & 	78.6 & 	71.7   \\
(Mbps) &	 & 	 & 	 & 	 & 	 &    \\
\hline
User &	 & 	 & 	 & 	 & 	 &    \\
Utility &34.5&	32.9&	34.0&	34.6&	34.3&	34.8   \\
\hline
\end{tabular}
\label{table:sim_1fBS1AP-noMacro}
\end{table*}

\subsection{A realistic scenario}
A suburban deployment scenario with a topology shown in Fig. \ref{fig_topology} is considered.
 A mBS is placed at the center
of the macrocell with radius 700m. Thirty (30)  mDevices are randomly
  dropped in the macrocell with uniform distribution. Due to the uniformly distributed mDevice locations,  some mDevices may be very close to fBSs.
We assume that  houses are located on 2D grid points
  with a center-to-center distance of 70m. Forty (40)  houses are randomly
  selected; each selected house is given one sDevice and one wDevice, which are randomly placed within the fBS coverage area with a
radius of 20m. In the use case of DBF, a WiFi AP is also placed in the selected house with a coverage radius of 20m. In this topology, the interference between houses is small, so each house is like an ``island". Therefore, the algorithm developed in the previous section is applicable to this topology. Although the traffic-balancing algorithm does not consider the interference among houses, other parts of our simulator (e.g., SINR evaluation) do.

\begin{figure}
  \center
  \includegraphics[width=3.1in]{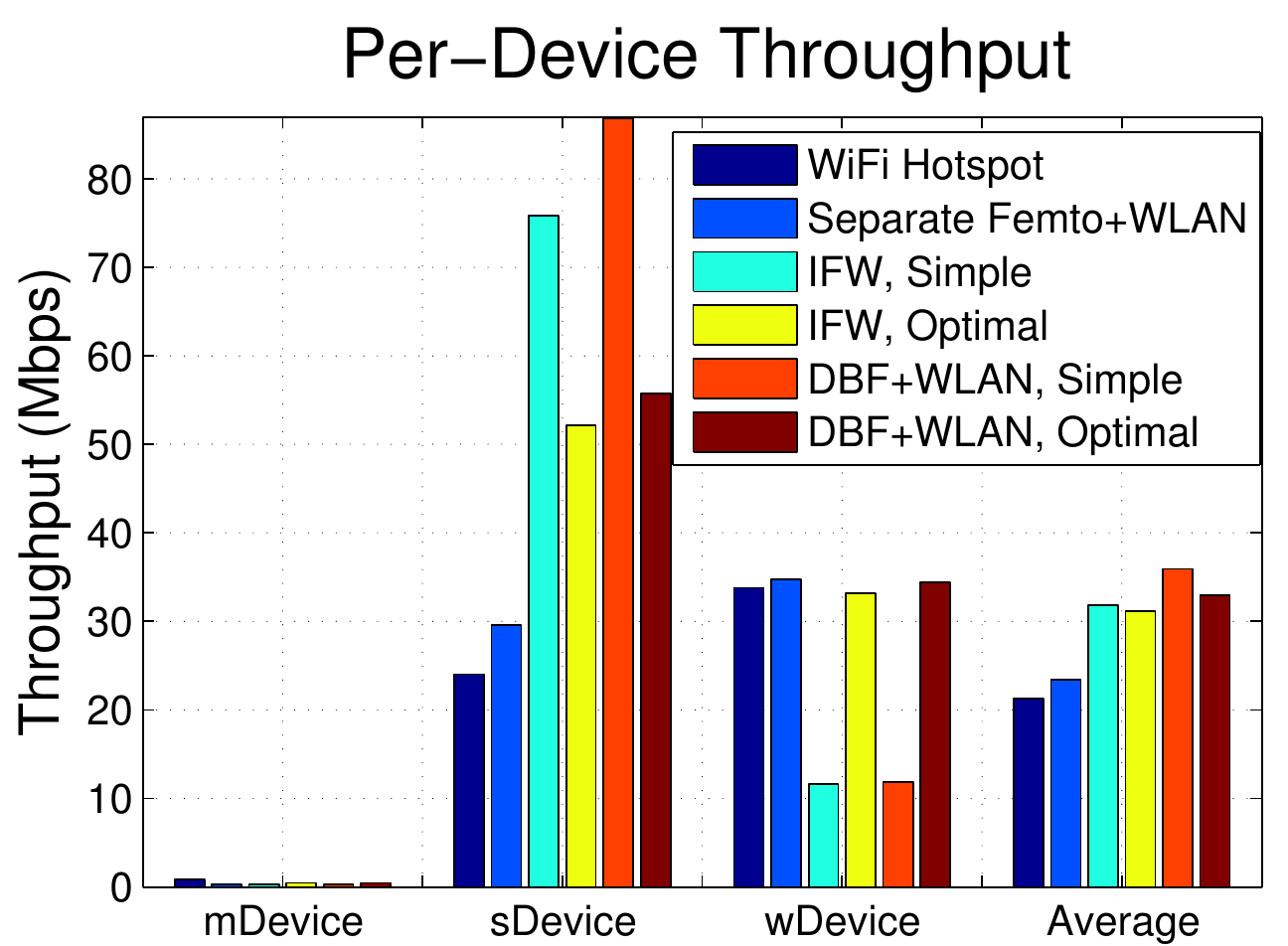}
  \caption{ Per-device throughput  when the aggregate wDevice load is
35Mbps (the highest WiFi physical layer rate is 72Mbps). The ``Average" metric is
averaged over all mDevices, sDevices and wDevices. } \label{fig:Thru-10MHz}
\end{figure}

\begin{figure*}[hpbt]
\centering \subfigure[]{
        \includegraphics[width=3.0in]{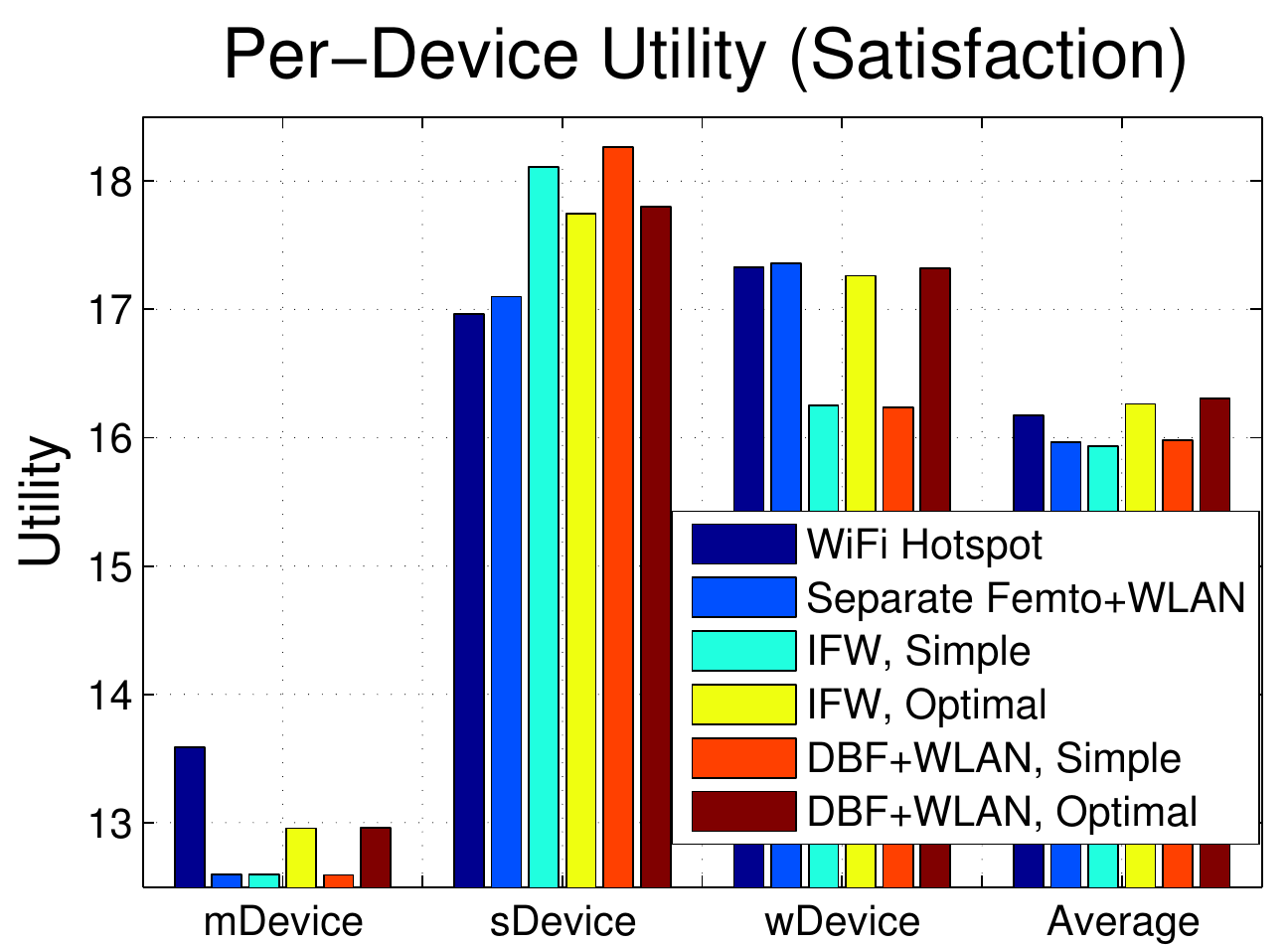}
        \label{fig:utility-10MHz-PerDevice}
        } \hfill
 \centering \subfigure[]{
        \includegraphics[width=3.0in]{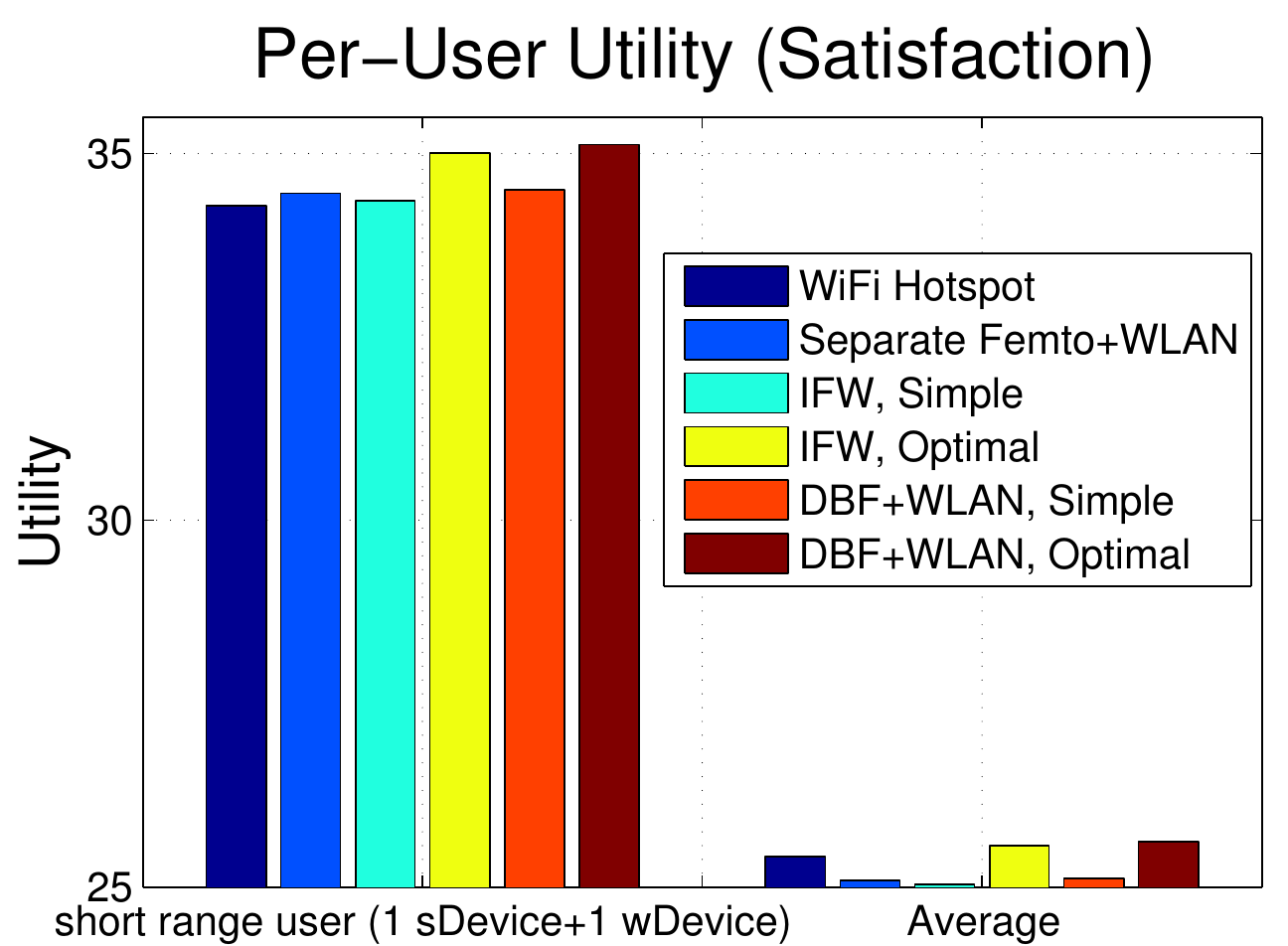}
        \label{fig:utility-10MHz-PerUser}
        }
\caption{ Per-device and per-user utilities  when the aggregate wDevice load is
35Mbps. In Fig. \ref{fig:utility-10MHz-PerDevice}, the ``Average" metric is
averaged over all mDevices, sDevices and wDevices; In Fig. \ref{fig:utility-10MHz-PerUser}, the ``Average" metric is
averaged over 30 macro users and 40 small cell users. } \label{fig:sim-10MHz}
\end{figure*}

We consider the scenario where the aggregate uplink and downlink wDevice load is 35Mbps (50\% of which is for uplink), which is
lower than the highest physical layer data rate 72Mbps for single-antenna 802.11n and lower than the highest achievable MAC layer throughput 61Mbps in our setup.
Figures \ref{fig:Thru-10MHz} and \ref{fig:sim-10MHz} show the throughput per device, utility per device, and utility per user. Note that each small cell user uses one sDevice and one wDevice; whereas each macro user uses one mDevice.  We have the
following observations from the figures. First, the small cell
optimal traffic-balancing algorithm significantly improves mDevice performance, while it does not significantly affect the
performance of sDevice or wDevice.  This can be verified by
comparing scenario ``IFW, Simple" with ``IFW, Optimal," and ``DBF+WLAN,
Simple" with ``DBF+WLAN, Optimal." The is mainly due to the licensed band
power control included in the algorithm which reduces the
interference from small cells to the macrocell.   Second, using two
bands  simultaneously  as in the IFW and DBF improves the average
utility and throughput of \emph{all devices}.  The proposed
traffic-balancing strategy shown in ``IFW, Optimal" and ``DBF+WLAN,
Optimal" obtain higher average utilities than the other cases,
including ``IFW, Simple" and ``DBF+WLAN, Simple." Third, there is very
little utility or throughput difference between the IFW and DBF
cases. Forth, the ``WiFi Hotspot" scenario has better good macro user utility,
which is mainly due to no interference to mDevice; in contrast, the fBSs in ``DBF+WLAN, Optimal" and ``IFW, Optimal" cases can only reduce, but not eliminate, interference to mDevices. While the proposed traffic-balancing algorithm does not
obtain the highest average throughput, this is not surprising, since
it is designed to achieve performance fairness through the utility
function. Note that in all cases discussed,  the total bandwidth is
30 MHz (licensed 10 MHz, unlicensed 20 MHz); however, the way this
bandwidth is shared among classes of devices is different in each
case.

\section{Discussions and Conclusion}\label{sec:conclusion}

Small cells have been considered as effective means to boost the
wireless capacity. In this paper, we have described the dual-band
femtocell (DBF) that simultaneously uses the LTE air interface in
licensed and unlicensed bands based on the LTE carrier aggregation
feature. We have proposed a channel access scheme for DBFs to access
the unlicensed band. Furthermore, we have described a traffic
balancing algorithm for small cells, including the DBF and the
Integrated Femto-WiFi (IFW) proposed by the Small Cell Forum, to  use
both licensed and unlicensed bands  in an optimized fashion thereby
improving the overall user utility/satisfaction from macrocell, small
cell and non-cellular WiFi-only devices. The algorithm searches for the
optimal power allocation in the licensed band, and the optimal
channel time usage in the unlicensed band. We have also proposed
practical algorithms to tune the unlicensed-band channel time usages
for IFWs and DBFs, which uses the WiFi and LTE air interfaces in the
unlicensed band, respectively.  Our results illustrate that, in
terms of average user utility, both IFW and DBF outperform current
WiFi hotspot and femtocell approaches  and thus are attractive
technologies for emerging small cell applications. While IFW and DBF
 have comparable performance, we have the following observations from an implementation
perspective.

\begin{itemize}
  \item \textbf{IFW:}
The IFW has separate cellular and WiFi radio interfaces in licensed
and unlicensed bands. Hence, it is backward compatible with existing
cellular and WLAN devices. However, special effort is required for a
single application flow to simultaneously use two radio interfaces
\cite{femto-forum-IFW}.

  \item \textbf{DBF:}
The DBF is actively under 3GPP LTE standardization
\cite{3GPP2014LTE-U},  where the unlicensed band is a secondary carrier in carrier aggregation.
Therefore, the DBF has a single radio interface. The DBF BS informs
its devices, via licensed-band control channel, about when and which
subchannel to receive their data in both licensed and unlicensed
bands. The DBF is backward compatible with existing LTE devices.
However, the DBF cannot serve WiFi-only devices. Therefore, it should be deployed or integrated with WiFi APs to serve WiFi-only devices.
\end{itemize}


\bibliographystyle{IEEEtran}

\begin{thebibliography}{10}

\bibitem{femto_Survey_JeffAndrews2008}
V.~Chandrasekhar, J.~Andrews, and A.~Gatherer, ``Femtocell networks: a
  survey,'' {\em IEEE Comm. Mag.}, pp.~59--67, Sep. 2008.

\bibitem{femto-PowerControl-2010Sundeep}
S.~Rangan, ``Femto-macro cellular interference control with subband scheduling
  and interference cancelation,'' {\em arXiv}, [Online]
  \url{http://arxiv.org/abs/1007.0507}, 2010.

\bibitem{femto-forum-IFW}
{Small Cell Forum}, ``{Integrated Femto-WiFi (IFW) networks},'' Whitepaper at
  \url{smallcellforum.org}, Feb. 2012.

\bibitem{feilu_femto_accessScheme}
F.~Liu, E.~Bala, E.~Erkip, and R.~Yang, ``A framework for femtocells to access
  both licensed and unlicensed bands,'' in {\em Proc. of the third
  International Workshop on Indoor and Outdoor Femto Cells (IOFC)}, Princeton,
  NJ, USA, May 13, 2011.

\bibitem{LTE-A}
{3GPP TS 36.300 v10.2.0}, ``{E-UTRA and E-UTRAN; Overall description; Stage 2
  (Release 10)},'' 2011.

\bibitem{3GPP2014LTE-U}
{Ericsson, Qualcomm, Huawei, Alcatel-Lucent}, ``{Study on licensed-assisted
  access using LTE},'' {\em RP-141664, 3GPP TSG RAN Meeting 65,}, Edinburgh,
  Scotland, 9-12 Sept. 2014.

\bibitem{feilu_trafficBalancing_conf}
F.~Liu, E.~Erkip, M.~Beluri, R.~Yang, and E.~Bala, ``Dual-band femtocell
  traffic balancing over licensed and unlicensed bands,'' in {\em Proc. of IEEE
  ICC}, Ottawa, ON, Canada, 10-15 June 2012.

\bibitem{unlicensed-band-LTE}
M.~Rahman, A.~Behravant, H.~Koorapaty, J.~Sachs, and K.~Balachandran,
  ``{License-exempt LTE systems for secondary spectrum usage: Scenarios and
  first assessment},'' in {\em Proc. of IEEE Symposium on New Frontiers in
  Dynamic Spectrum Access Networks (DySPAN)}, May 2011.

\bibitem{huawei_unlic_LTE}
L.~Sun, ``The unlicensed spectrum usage for future {IMT} technologies,'' in
  {\em Proc. of The 6th International Workshop - Vision and Technology Trends
  for 5G}, Seoul, Korea, Sept. 04, 2013.

\bibitem{trafficBalance_qos_mag}
M.~Bennis, M.~Simsek, A.~Czylwik, W.~Saad, S.~Valentin, and M.~Debbah, ``{When
  cellular meets WiFi in wireless small cell networks},'' {\em {IEEE
  communications magazine}}, vol.~51, Jun. 2013.

\bibitem{fujitsu_trafficBalance_IFW_IncludewifiOnly}
A.~Elsherif, W.-P. Chen, A.~Ito, and Z.~Ding, ``Adaptive small cell access of
  licensed and unlicensed bands,'' [Online]
  \url{http://www.fujitsu.com/downloads/SVC/fla/research/Adaptive-Small-Cell-Access-of-Licensed-and-Unlicensed-Bands.pdf}
  , 2013.

\bibitem{ofdm_multiuser}
N.~Ksairi, P.~Bianchi, and P.~Ciblat, ``Nearly optimal resource allocation for
  downlink ofdma in {2-D} cellular networks,'' {\em {IEEE Trans. On Wireless
  Comm.}}, vol.~10, pp.~2101--2115, July 2011.

\bibitem{802.11-2007}
{IEEE Std 802.11-2007 (Revision of Std 802.11-1999)}, ``{Part II: Wireless LAN
  MAC and PHY Specifications},'' 2007.

\bibitem{log_utility}
J.~Mo and J.Walrand, ``Fair end-to-end window-based congestion control,'' {\em
  {IEEE Trans. Netw.}}, pp.~556--567, Oct. 2000.

\bibitem{queuing-theory-textbook}
D.~Bertsekas and R.~Gallager, {\em Data Networks}.
\newblock {Prentice Hall}, 1992.

\bibitem{Bianchi}
G.~Bianchi, ``{Performance analysis of the IEEE 802.11 distributed coordination
  function},'' {\em {IEEE Journal on Selected Areas in Communications}},
  vol.~18, pp.~535--547, Mar. 2000.

\bibitem{WiFi_modeling_Foh_freezingCounter}
C.~H. Foh and J.~W. Tantra, ``{comments on IEEE 802.11 Saturation Throughput
  Analysis with Freezing of Backoff Counters},'' {\em {IEEE Communications
  Letters}}, vol.~9, pp.~130--132, Feb. 2005.

\bibitem{WiFi_analysis_freezingCounter_Lee2005Conf}
Y.~Lee, D.~H. Han, and C.~G. Park, ``{IEEE 802.11 saturation throughput
  analysis with freezing of backoff counters},'' in {\em Proc. of ICCOM'05},
  Stevens Point, Wisconsin, USA, 2005.

\bibitem{ConvexOptimization_textbook}
S.~P. Boyd and L.~Vandenberghe, {\em Convex Optimization}.
\newblock Cambridge Univ. Press, 2004.

\bibitem{cap-limited-water-filling-wcnc2009}
K.~Son, B.~C. Jung, S.~Chong, and D.~K. Sung, ``Opportunistic underlay
  transmission in multi-carrier cognitive radio systems,'' in {\em Proc. of
  WCNC 2009}.

\bibitem{cap-limited-water-filling-EURASIP2008}
N.~Papandreou and T.~Antonakopoulos, ``Bit and power allocation in constrained
  multicarrier systems: The single-user case,'' {\em EURASIP JNL on Advances in
  Signal Processing}, Article ID 643081, Oct. 2008.

\bibitem{LteRateApprox}
P.~Mogensen, W.~Na, I.~Kovacs, F.~Frederiksen, A.~Pokhariyal, K.~Pedersen,
  T.~Kolding, K.~Hugl, and M.~Kuusela, ``{LTE} capacity compared to the shannon
  bound,'' in {\em Proc. of Vehicular Technology Conference (VTC2007-Spring)},
  April 2007.

\bibitem{3GPP_TR_36.921_femto}
{3GPP TR 36.921 v10.0.0}, ``{E-UTRA; FDD} home {eNode B (HeNB)} radio frequency
  {(RF)} requirements analysis,'' 2011.

\bibitem{mBS_broadcastLocRB}
A.~Adhikary, V.~Ntranos, and G.~Caire, ``Cognitive femtocells: Breaking the
  spatial reuse barrier of cellular systems,'' in {\em {Proc. of Information
  Theory and Applications Workshop (ITA)}}, Feb. 2011.

\bibitem{RF-fingerprint}
W.~C. Suski, M.~A. Temple, M.~J. Mendenhall, and R.~F. Mills, ``Radio frequency
  fingerprinting commercial communication devices to enhance electronic
  security,'' {\em Int'l Journal of Electronic Security and Digital Forensics},
  pp.~301--322, Oct 2008.

\bibitem{solve_WiFi_Anomaly_letter_AdjustW}
H.~Kim, S.~Yun, I.~Kang, and S.~Bahk, ``{Resolving 802.11 performance anomalies
  through QoS differentiation},'' {\em IEEE Communications Letters},
  pp.~655--657, July 2005.

\bibitem{Computing-textbook-bisection}
W.~H. Press, B.~P. Flannery, S.~A. Teukolsky, and W.~T. Vetterling, {\em
  {Numerical recipes in Fortran: the art of scientific computing, 2nd ed.}}
\newblock {Cambridge University Press}, 1992.

\bibitem{equal_access_measure_infocom2003}
M.~Heusse, F.~Rousseau, G.~Berger-Sabbatel, and A.~Duda, ``{Performance anomaly
  of 802.11b},'' in {\em {Proc. of INFOCOM}}, 2003.

\bibitem{CoopMAC_WirelessMag2006}
P.~Liu, Z.~Tao, Z.~Lin, E.~Erkip, and S.~S. Panwar, ``Cooperative wireless
  communications: a cross layer approach,'' {\em IEEE Wireless Communications},
  vol.~13, pp.~84--92, Aug. 2006.

\bibitem{EqualAccess_CoopMAC_Jsac2007}
P.~Liu, Z.~Tao, S.~Narayanan, T.~Korakis, and S.~S. Panwar, ``{CoopMAC: a
  cooperative MAC for wireless LANs},'' {\em {IEEE Journal on Selected Areas in
  Communications}}, vol.~25, pp.~340--354, Feb. 2007.

\bibitem{femto-forum-whitepaper}
{Small Cell Forum}, ``Interference management in {OFDMA} femtocells,''
  Whitepaper at \url{smallcellforum.org}, Mar. 2010.

\bibitem{802.11n}
{IEEE Std 802.11n-2009}, ``{Part 11: Wireless LAN MAC and PHY
  Specifications},'' 2009.

\end{thebibliography}

\begin{IEEEbiography}[{\includegraphics[width=1in,height=1.25in,clip,keepaspectratio]{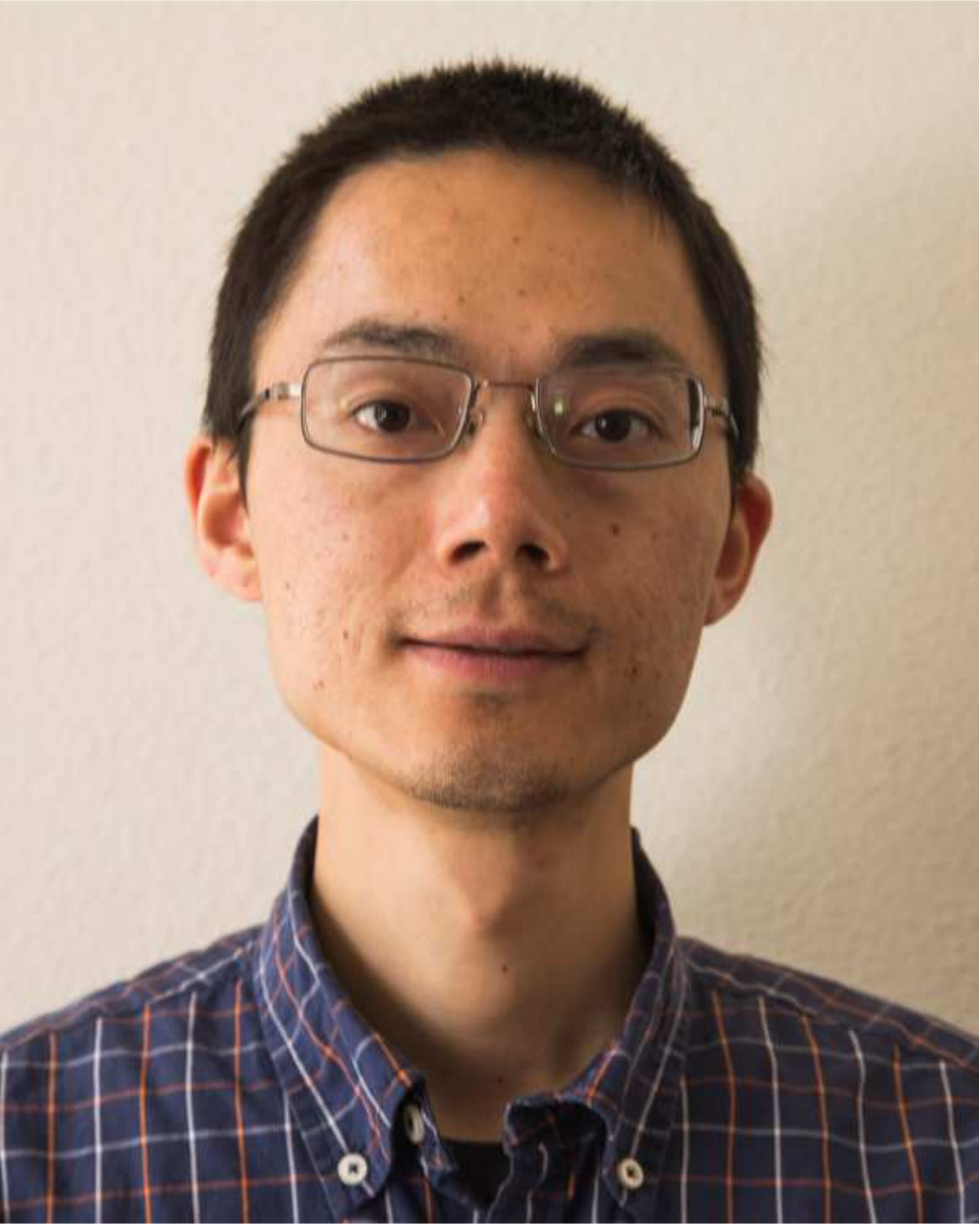}}]{Feilu Liu}
Feilu Liu received the B.E. degree in electronic engineering from Southeast University, Nanjing, China, and the M.S. and Ph.D. degrees in electrical engineering from Polytechnic school of engineering, New York University. He has been with Qualcomm Technologies, CA as a systems engineer since 2012. His previous work experience includes software engineer at Alcatel-Lucent, China and intern at InterDigital, NY. He has been working on the systems design for LTE MAC/RLC/PDCP/RRC layer implementations. He is interested in wireless networking protocol and algorithm design.
\end{IEEEbiography}

\begin{IEEEbiography}[{\includegraphics[width=1in,height=1.25in,clip,keepaspectratio]{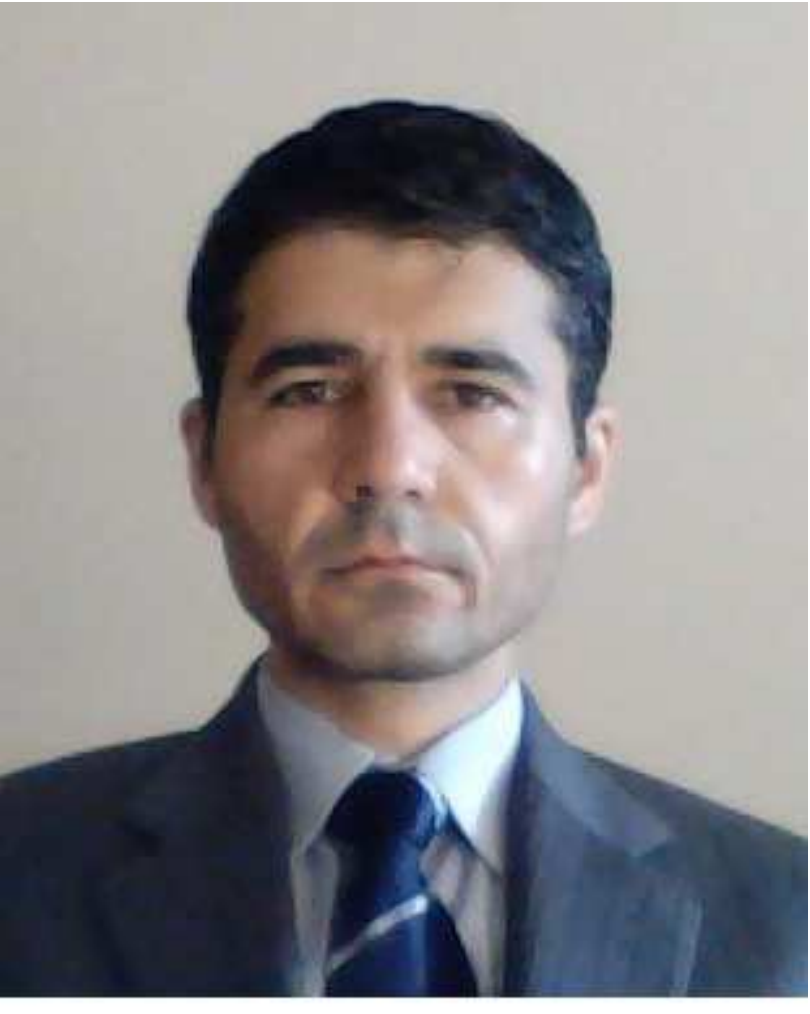}}]{Erdem Bala}
Erdem Bala got his BSc and MSc degrees from Bogazici University, Istanbul, Turkey and his PhD degree from the University of Delaware, DE, all in electrical engineering. He has been with InterDigital, NY as a research engineer since 2007. His previous work experience includes positions as R\&D engineer at Nortel Networks and intern at Mitsubishi Research Labs. At InterDigital, he has worked on the standardization of 3GPP LTE and LTE-Advanced, advanced relaying schemes, coexistence in unlicensed spectrum, and waveform design. Currently, he is involved in the design of 5G air interface for future wireless communication systems.
\end{IEEEbiography}

\begin{IEEEbiography}[{\includegraphics[width=1in,height=1.25in,clip,keepaspectratio]{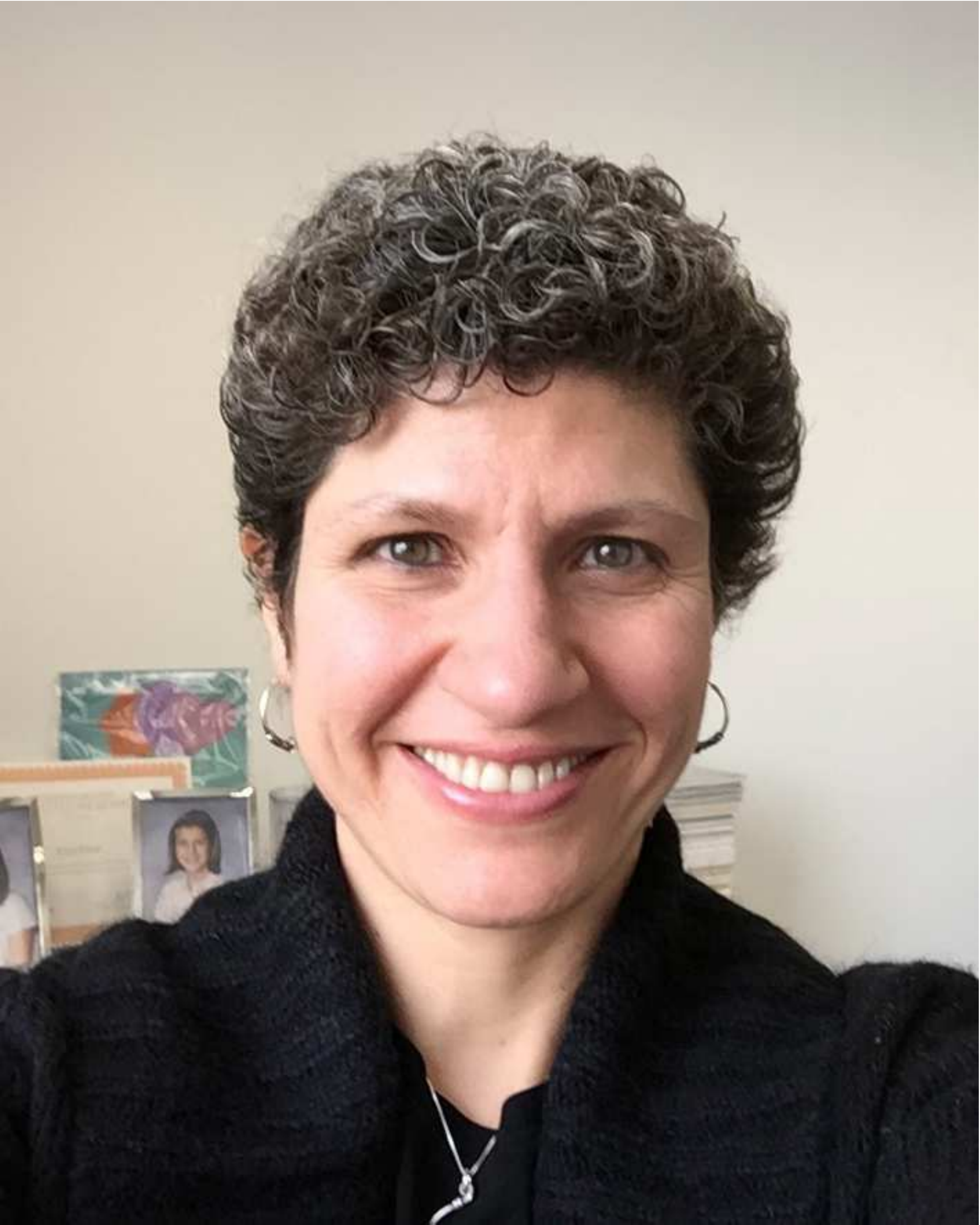}}]{Elza Erkip}
Elza Erkip (S'93-M'96-SM'05-F'11) received the B.S. degree in electrical and electronics engineering from Middle East Technical University, Ankara, Turkey, and the M.S. and Ph.D. degrees in electrical engineering from Stanford University, Stanford, CA, USA. Currently, she is a Professor of electrical and computer engineering with New York University Polytechnic School of Engineering, Brooklyn, NY, USA. Her research interests are in information theory, communication theory, and wireless communications.

Dr. Erkip is a member of the Science Academy Society of Turkey and is among the Thomson Reuters 2014 Edition of Highly Cited Researchers. She received the NSF CAREER award in 2001, the IEEE Communications Society Stephen O. Rice Paper Prize in 2004, the IEEE ICC Communication Theory Symposium Best Paper Award in 2007, and the IEEE Communications Society Award for Advances in Communication in 2013. She co-authored a paper that received the IEEE International Symposium on Information Theory Student Paper Award in 2007. Currently, she is a Member of the Board of Governors of the IEEE Information Theory Society and a Guest Editor of the IEEE JOURNAL ON SELECTED AREAS IN COMMUNICATIONS. Dr. Erkip was a Distinguished Lecturer of the IEEE Information Theory Society from 2013 to 2014, an Associate Editor of the IEEE TRANSACTIONS ON INFORMATION THEORY from 2009 to 2011, an Associate Editor of the IEEE TRANSACTIONS ON COMMUNICATIONS from 2006 to 2008, a Publications Editor of the IEEE TRANSACTIONS ON INFORMATION THEORY from 2006 to 2008 and a Guest Editor of the IEEE SIGNAL PROCESSING MAGAZINE in 2007. She was a General Chair for the IEEE International Symposium of Information Theory in 2013, a Technical Program Chair for the International Symposium on Modeling and Optimization in Mobile, Ad Hoc, and Wireless Networks (WiOpt) in 2011, a Technical Program Chair for the IEEE GLOBECOM Communication Theory Symposium in 2009, the Publications Chair for the IEEE Information Theory Workshop, Taormina, in 2009, the Technical Area Chair for the MIMO Communications and Signal Processing track of Asilomar Conference on Signals, Systems, and Computers in 2007, and a Technical Program Chair for the IEEE Communication Theory Workshop in 2006.
\end{IEEEbiography}

\begin{IEEEbiography}[{\includegraphics[width=1in,height=1.25in,clip,keepaspectratio]{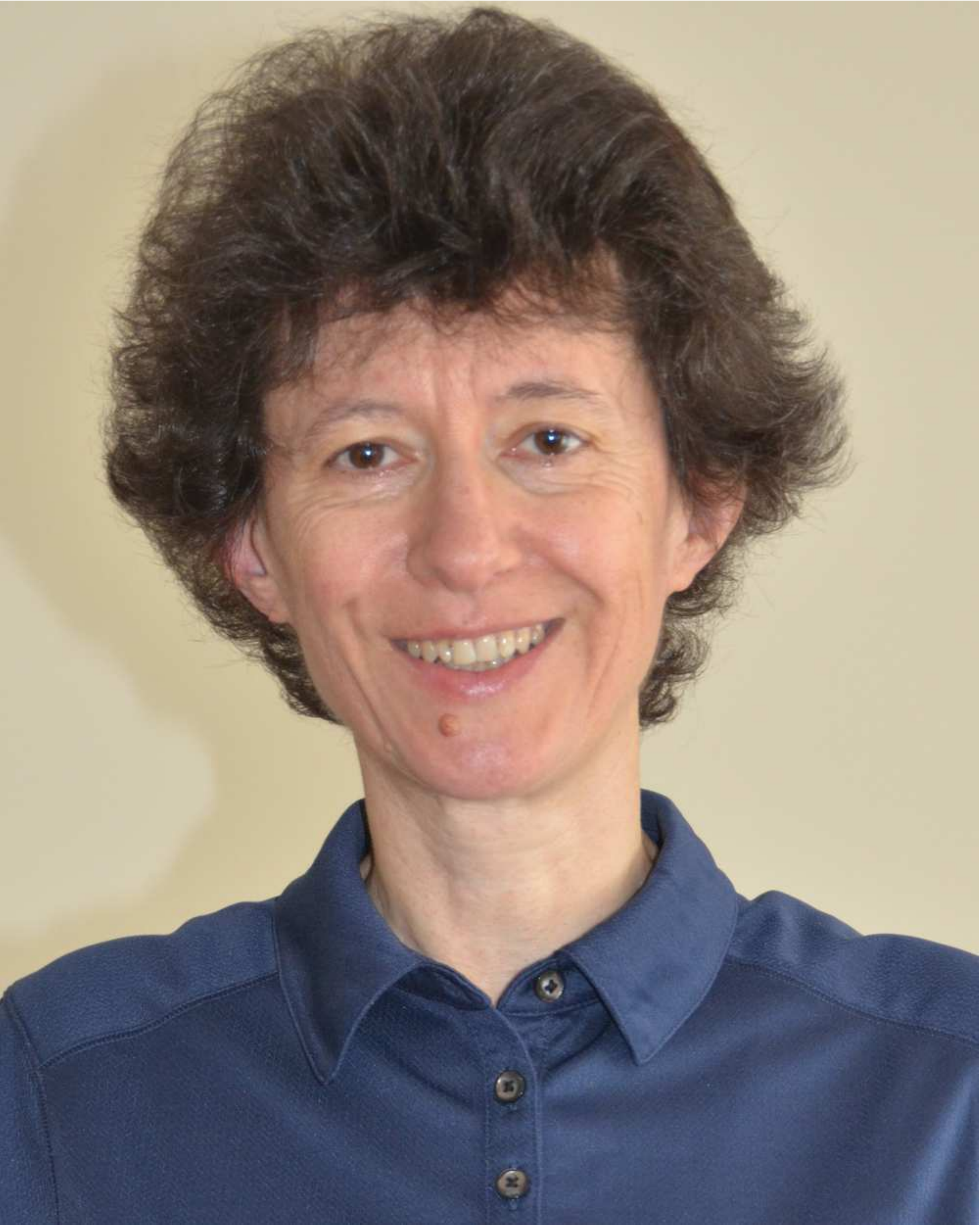}}]{Mihaela Beluri}
Mihaela Beluri (M'90) received the M.S. degree in electrical engineering from the Polytechnic University of Bucharest, Romania. She is currently a Principal Engineer with InterDigital Communications, Melville, NY, working on millimeter wave technologies. Her recent experience includes dynamic spectrum management and shared spectrum technologies, algorithm design, modeling and simulations for WCDMA, HSPA and LTE systems. She has authored or coauthored IEEE conference publications and holds several patents.
\end{IEEEbiography}

\begin{IEEEbiography}[{\includegraphics[width=1in,height=1.25in,clip,keepaspectratio]{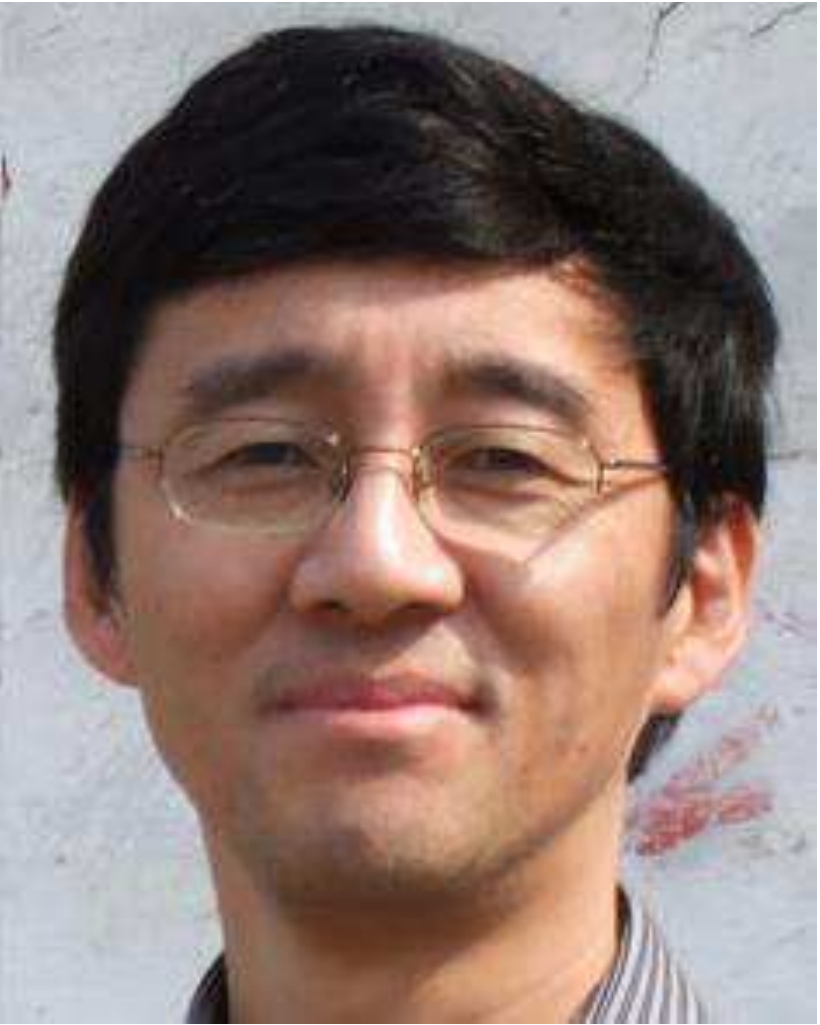}}]{Rui Yang}
Rui Yang received the M.S. and Ph.D. degrees in electrical engineering from the University of Maryland, College Park, in 1987 and 1992, respectively. He has 15 years of experience in the research and development of wireless communication systems. Since he joined InterDigital Communications in 2000, he has led several product development and research projects. He is currently a Principle Engineer at InterDigital Labs, leading a project on baseband and RF waveforms for future wireless communication systems. His interests include digital signal processing and air interface design. He has received more than 15 patent awards in those areas.
\end{IEEEbiography}

\end{document}